\documentclass[10pt,conference,compsocconf]{IEEEtran}
\IEEEoverridecommandlockouts
\usepackage{cite}
\usepackage{amsmath,amssymb,amsfonts}
\usepackage{algorithmic}
\usepackage{graphicx}
\usepackage{textcomp}
\usepackage{xcolor}
\usepackage{subscript}
\usepackage{balance}

\usepackage{url}
\usepackage{booktabs}
\usepackage{caption}
\usepackage{enumitem}
\usepackage{dsfont}
\usepackage{caption}
\usepackage{subcaption}
\usepackage{xspace}
\usepackage{amsthm}
\usepackage{multirow}
\theoremstyle{definition}

\usepackage[ruled, vlined, linesnumbered]{algorithm2e} 

\usepackage{pifont}
\DeclareMathOperator*{\argmax}{arg\,max}
\DeclareMathOperator*{\argmin}{arg\,min}

\def\BibTeX{{\rm B\kern-.05em{\sc i\kern-.025em b}\kern-.08em
    T\kern-.1667em\lower.7ex\hbox{E}\kern-.125emX}}
\begin{document}

\setlength{\textfloatsep}{0.11cm}
\setlength{\dbltextfloatsep}{0.11cm}
\setlength{\abovecaptionskip}{0.11cm}
\setlength{\skip\footins}{0.11cm}

\title{Set2Box: Similarity Preserving Representation Learning for Sets}

\author{
	\IEEEauthorblockN{Geon Lee}
	\IEEEauthorblockA{KAIST AI\\geonlee0325@kaist.ac.kr}
	\and
	\IEEEauthorblockN{Chanyoung Park}
	\IEEEauthorblockA{KAIST ISysE \& AI\\cy.park@kaist.ac.kr}
	\and
	\IEEEauthorblockN{Kijung Shin}
	\IEEEauthorblockA{KAIST AI \& EE\\kijungs@kaist.ac.kr}
}

\maketitle
    
\newcommand\red[1]{\textcolor{red}{#1}}
\newcommand\blue[1]{\textcolor{blue}{#1}}
\newcommand\geon[1]{\textcolor{blue}{[Geon:#1]}}

\newcommand{\smallsection}[1]{{\noindent {\bf{\underline{\smash{#1}}}}}}
\newtheorem{obs}{\textbf{Observation}}
\newtheorem{defn}{\textbf{Definition}}
\newtheorem{thm}{\textbf{Theorem}}
\newtheorem{axm}{\textbf{Axiom}}
\newtheorem{lma}{\textbf{Lemma}}
\newtheorem{prb}{\textbf{Problem}}

\newcommand{\cmark}{\ding{51}}%
\newcommand{\xmark}{\ding{55}}%

\newcommand{\method}{\textsc{Set2Box}\xspace}
\newcommand{\methodq}{\textsc{Set2Box}\textsuperscript{+}\xspace}
\newcommand{\methodpq}{\textsc{Set2Box-PQ}\xspace}
\newcommand{\methodfeat}{\textsc{Set2Box-feat}\xspace}
\newcommand{\methodbq}{\textsc{Set2Box-BQ}\xspace}
\newcommand{\methodo}{\textsc{Set2Box-order}\xspace}
\newcommand{\rh}{\textsc{Set2Bin}\xspace}
\newcommand{\vect}{\textsc{Set2Vec}\xspace}
\newcommand{\vectmlp}{\textsc{Set2Vec}\textsuperscript{+}\xspace}
\newcommand{\vectds}{\textsc{Set2Vec-DS}\xspace}

\definecolor{myred}{HTML}{D62728}
\definecolor{mygreen}{HTML}{2CA02C}
\definecolor{myblue}{HTML}{1F77B4}
\definecolor{mypurple}{HTML}{9467BD}
\definecolor{myorange}{HTML}{FF7F0E}
\definecolor{myolive}{HTML}{BCBD22}
\definecolor{mybrown}{HTML}{8C564B}

\newcommand*{\medcap}{\mathbin{\scalebox{1.2}{\ensuremath{\bigcap}}}}
\newcommand*{\medcup}{\mathbin{\scalebox{1.2}{\ensuremath{\bigcup}}}}

\def\lc{\left\lceil}   
\def\rc{\right\rceil}

	\begin{abstract}
		Sets have been used for modeling various types of objects (e.g., a document as the set of keywords in it and a customer as the set of the items that she has purchased).
Measuring similarity (e.g., Jaccard Index) between sets has been a key building block of a wide range of applications, including, plagiarism detection, recommendation, and graph compression.
However, as sets have grown in numbers and sizes, the computational cost and storage required for set similarity computation have become substantial, and this has led to the development of hashing and sketching based solutions.



In this work, we propose \method, a learning-based approach for compressed representations of sets from which various similarity measures can be estimated accurately in constant time.
The key idea is to represent sets as boxes
to precisely capture overlaps of sets. 
Additionally, based on the proposed box quantization scheme, we design \methodq, which yields more concise but more accurate box representations of sets.
Through extensive experiments on $8$ real-world datasets, 
we show that, compared to baseline approaches, \methodq is \textit{(a) Accurate:} achieving up to \textit{40.8$\times$ smaller  estimation error} while requiring 60\% fewer bits to encode sets, \textit{(b) Concise:} yielding up to \textit{96.8$\times$ more concise} representations with similar estimation error, and \textbf{(c) Versatile:} enabling the estimation of four set-similarity measures from a single representation of each set.
	\end{abstract}
	

	\section{Introduction}
	\label{sec:intro}

Sets are ubiquitous, modeling various types of objects in many domains, including \textbf{(a) a document:} modeled as the set of keywords in it, \textbf{(b) a customer:} modeled as the set of the items that she has purchased, \textbf{(c) a social circle:} modeled as the set of its members, and \textbf{(d) a question on online Q/A platforms:} modeled as the set of tags attached to the question. Moreover,
a number of set similarity measures (e.g., Jaccard Index and Dice Index), most of which are based on the overlaps between sets, have been developed.

As a result of the omnipresence of sets, measuring their similarity has been employed as a fundamental building block of a wide range of applications, including the following:

\noindent $\circ$ \textbf{Plagiarism Detection:}
Plagiarism is a critical problem in the digital age, where a vast amount of resources is accessible. A text is modeled as a ``bag of words,'' and texts whose set representations are highly similar are suspected of plagiarism~\cite{moussiades2005pdetect}. 

\noindent $\circ$ \textbf{Gene Expression Mining:} Mining gene expressions is useful for understanding clinical conditions (e.g., tumor and cancer). 
The functionality of a set of genes is estimated by comparing the set with other sets with known functionality~\cite{yousri2011associating}.

\noindent $\circ$ \textbf{Recommendation:}
Recommendation 
is essential to support users in finding relevant items.
To this end, it is useful to identify users with similar tastes (e.g., users who purchased a similar set of items and users with similar activities) \cite{koren2008factorization,guerraoui2020smaller}.

\noindent $\circ$ \textbf{Graph Compression:}
As large-scale graphs are omnipresent, compressing them into coarse-grained summary graphs so that they fit in main memory is important. In many graph compression algorithms, nodes with similar sets of neighbors are merged into a supernode to yield a concise summary graph while minimizing the information loss~\cite{shin2019sweg,navlakha2008graph}.

\noindent $\circ$ \textbf{Medical Image Analysis:}
CT or MRI provide exquisite details of inner body (e.g., brain), and they are often described as a collection of spatially localized anatomical features termed ``keypoints''.
Sets of keypoints from different images are compared to diagnose and investigate diseases~\cite{chauvin2020neuroimage,chauvin2021efficient,toews2010feature}.

As sets grow in numbers and sizes, computation of set similarity requires substantial computational cost and storage.
For example, similarities between tens of millions of nodes, which are represented as neighborhood sets of up to millions of neighbors, were measured for graph compression \cite{shin2019sweg}.
Moreover, similarities between tens of thousands of movies, which are represented as sets of up to hundreds of thousands of users who have rated them, were measured for movie recommendation \cite{koren2008factorization}.


In order to reduce the space and computation required for set-similarity computation, a number of approaches based on hashing and sketching~\cite{guerraoui2020smaller,cui2005exploring} have been developed.
While their simplicity and theoretical guarantees are tempting, significant gains are expected if patterns in a given collection of sets can be learned and exploited.

In this paper, we propose \method, a learning-based approach for compressed representations of sets from which various similarity measures can be estimated accurately in constant time.  
The key idea of \method is to represent sets as boxes to accurately capture the overlaps between sets and thus their similarity based on them.
Specifically, by utilizing the volumes of the boxes to approximate the sizes of the sets, \method derives representations that are: \textbf{(a) Concise:} can represent sets of arbitrary sizes using the same number of bits, \textbf{(b) Accurate:} can accurately model overlaps between sets, and \textbf{(c) Versatile:} can be used to estimate various set similarity measures in a constant time.
These properties are supported by the 
geometric nature of boxes, which share primary characteristics of sets.
In addition, we propose \methodq, which yields even more concise but more accurate boxes based on the proposed box quantization scheme. 
We summarize our contributions as follows:

\begin{itemize}[leftmargin=*]
    \item \textbf{Accurate \& Versatile Algorithm:} We propose \method, a set representation learning method that accurately preserves similarity between sets in terms of four measures. 
    \item \textbf{Concise \& Effective Algorithm:} We devise \methodq to enhance \method through an end-to-end box quantization scheme. It yields up to $40.8\times$ more accurate similarity estimation while requiring $60\%$ fewer bits than its competitors.
    \item \textbf{Extensive Experiments:} Using 8 real-world datasets, we validate the advantages of \methodq over its competitors and the effectiveness of each of its components.
\end{itemize}
For \textbf{reproducibility}, the code and data are available at \url{https://github.com/geon0325/Set2Box}.


\begin{table}[t!]
	\begin{center}
		\caption{\label{tab:notations}Frequently-used symbols.}
		\scalebox{0.985}{
			\begin{tabular}{c|l}
				\toprule
				\textbf{Notation} & \textbf{Definition}\\
				\midrule
				$\mathcal{S}=\{s_1,...,s_{|\mathcal{S}|}\}$ & set of sets\\
                $\mathcal{E}=\{e_1,...,e_{|\mathcal{E}|}\}$ & set of entities\\
                \midrule
                $\mathrm{B} = (\mathrm{c}, \mathrm{f})$ & a box with center $\mathrm{c}$ and offset $\mathrm{f}$\\
                $\mathbb{V}(\mathrm{B})$ & volume of box $\mathrm{B}$\\
                \midrule
				$\mathcal{T}^{+}$ and $\mathcal{T}^{-}$ & a set of positive \& negative samples\\
				\midrule
				$\mathbf{Q}^{\mathrm{c}}\in \mathbb{R}^{|\mathcal{E}|\times d}$ & center embedding matrix of entities\\
				$\mathbf{Q}^{\mathrm{f}}\in \mathbb{R}_{+}^{|\mathcal{E}|\times d}$ & offset embedding matrix of entities\\
				\midrule
				$D$ & number of subspaces\\
				$K$ & number of key boxes in each subspace\\
				\bottomrule
			\end{tabular}}
	\end{center}
\end{table}

In Section~\ref{sec:related}, we review related work.
In Section~\ref{sec:prelim}, we define the problem of similarity-preserving set embedding and discuss intuitive approaches.
In Section~\ref{sec:method}, we present 
\method and \methodq. 
In Section~\ref{sec:experiments}, we provide experimental results.
In Section~\ref{sec:discussions}, we analyze the considered methods.
Lastly, we offer conclusions in Section~\ref{sec:summary}.
	
	\section{Related Work}
	\label{sec:related}
	Here, we review previous studies related to our work.

\smallsection{Similarity-Preserving Embedding:}
Representation learning for preserving similarities between instances has been studied for graphs~\cite{jin2021node,xie2019sim2vec,tsitsulin2018verse,ou2016asymmetric}, images~\cite{tung2019similarity,li2017deep,zhu2016deep}, and texts~\cite{li2021berttocnn}.
These methods aim to yield high-quality embeddings by minimizing the information loss of the original data.
However, most of them are designed to preserve the predetermined similarity matrix, which are not extensible to new measures~\cite{tsitsulin2018verse,ou2016asymmetric}.
In this paper, we focus on the problem of learning similarity-preserving representations for sets, and we aim to learn a versatile representation of sets, which various similarity measures (e.g., Jaccard Index and Dice Index) can be estimated from.

\smallsection{Box Embedding:} 
Boxes~\cite{vilnis2018probabilistic} are useful abstractions to express high-order information of the data.
Thanks to their powerful expressiveness, they have been used in diverse applications including knowledge bases~\cite{abboud2020boxe,liu2021neural,onoe2021modeling,chen2021probabilistic,ren2019query2box,patel2020representing}, word embedding~\cite{dasgupta2021word2box}, image embedding~\cite{rau2020predicting}, and recommender systems~\cite{zhang2021learning,deng2021box4rec}.
For instance, Query2Box~\cite{ren2019query2box} uses boxes to embed queries with conjunctions ($\wedge$) or logical disjunctions ($\lor$).
Zhang et al.~\cite{zhang2021learning} represent users as boxes to accurately model the users' preferences to the items. 
However, in this work, we embed sets as boxes to accurately preserve their structural relationships and also similarities between them.
In an algorithmic aspect, methods for improving the optimization of learning boxes have been presented, and examples include smoothing hard edges using Gaussian convolutions~\cite{li2018smoothing} and improving the parameter identifiability of boxes using Gumbel random variables~\cite{dasgupta2020improving}.

\smallsection{Set Embedding:}
The problem of embedding sets has attracted much attention, with unique requirements of permutation invariance and size flexibility.
For example, DeepSets~\cite{zaheer2017deep} uses simple symmetric functions over input features, and Set2Set~\cite{vinyals2015order} is based on a LSTM-based pooling function.
Set Transformer~\cite{lee2019set} uses an attention-based pooling function to aggregate information of the entities.
Despite their promising results in some predictive tasks, they suffer from several limitations.
First, they require attribute information of entities, which in fact largely affects the quality of the set embeddings.
In addition, set representations are trained specifically for downstream tasks, and thus they may lose explicit similarity information of sets, which we aim to preserve in this paper.
In another aspect, sets can be represented as compact binary vectors by hashing or sketching~\cite{guerraoui2020smaller,cui2005exploring}, without requiring attribute information. Such binary vectors are used by 
Locality Sensitive Hashing (LSH) and its variants~\cite{indyk1998approximate,charikar2002similarity,ji2012super}
for a rapid search of similar sets based on a predefined similarity measure (e.g., Jaccard Index).
Refer to Section~\ref{sec:prelim} for further discussion of set embedding methods.



\smallsection{Differentiable Product Quantization:}
Product quantization \cite{jegou2010product,ge2013optimized} is an effective strategy for vector compression.
Recently, deep learning methods for learning discrete codes in an end-to-end manner have been proposed~\cite{chen2020differentiable,chen2018learning}, and they have been applied in knowledge graphs~\cite{sachan2020knowledge} and image retrieval~\cite{klein2019end,jang2021self,morozov2019unsupervised}.
In this paper, we propose a novel box quantization method for compressing boxes while preserving their original geometric properties.
	
	\section{Preliminaries}
	\label{sec:prelim}
	In this section, we introduce notations and define the problem. 
Then, we review some intuitive methods for the problem.

\smallsection{Notations:}
Consider a set $\mathcal{S}=\{s_1,\cdots,s_{|\mathcal{S}|}\}$ of \textit{sets} and a set $\mathcal{E}=\{e_1,\cdots, e_{|\mathcal{E}|}\}$ of \textit{entities}.
Each set $s\in \mathcal{S}$ is a non-empty subset of $\mathcal{E}$ and its size (i.e., cardinality) is denoted by $|s|$.
A representation of the set $s$ is denoted by $\mathrm{z}_s$ and its encoding cost (the number of bits to encode $\mathrm{z}_s$) in bits is denoted by $Cost(\mathrm{z}_s)$.
Refer to Table~\ref{tab:notations} for frequently-used notations. 

\smallsection{Problem Definition:}
The problem of learning similarity-preserving set representations, which we focus in this work, is formulated as:

\begin{prb}[Similarity-Preserving Set Embedding] \ \label{defn:problem}
\begin{itemize}[leftmargin=*]
    \item \textbf{Given:} \textit{(1) a set $\mathcal{S}$ of sets and (2) a budget $b$} 
    \item \textbf{Find:} \textit{a latent representation $\mathrm{z}_s$ of each set $s\in\mathcal{S}$}
    \item \textbf{to Minimize:} \textit{the difference between (1) the similarity between $s$ and $s'$, and  (2) the similarity between $\mathrm{z}_s$ and  $\mathrm{z}_{s'}$ for all $s\neq s'\in \mathcal{S}$}
    \item \textbf{Subject to:} \textit{the total encoding cost $\textit{Cost}(\{\mathrm{z}_s:s\in \mathcal{S}\}) \leq b$.}
\end{itemize}
\end{prb}
\noindent In this paper, we consider four set-similarity measures and use the mean squared error (MSE)\footnote{$\sum_{s\neq s'\in\mathcal{S}} {|\text{sim}(s,s') - \widehat{\text{sim}}(\mathrm{z}_{s}, \mathrm{z}_{s'})|}^2$  $\text{sim}(\cdot,\cdot)$ and $\widehat{\text{sim}}(\cdot,\cdot)$ are similarity between sets and that between latent representations, respectively.} to measure the differences, while our proposed methods are not specialized to the choices.


\smallsection{Desirable Properties:}
We expect 
set embeddings for Problem~\ref{defn:problem} to have the following desirable properties: 

\begin{itemize}[leftmargin=*]
    \item \textbf{Accuracy:} How can we accurately preserve similarities between  sets? Similarities approximated using learned representations should be close to ground-truth similarities.
    \item \textbf{Conciseness:} How can we obtain compact representations that give a good trade-off between accuracy and encoding cost? It is desirable to use less amount of memory to store embeddings while keeping them informative.
    \item \textbf{Generalizability:} Due to the size flexibility of sets, there are infinitely many number of combinations of entities, and thus retraining the entire model for new sets is intractable. It is desirable for a model to be generalizable to unseen sets.
    \item \textbf{Versatility:} While there have been various definitions of set similarities, the choice of the similarity metric plays a key role in practical analyses and applications. This motivates us to learn versatile representations of sets that can be used to approximate diverse similarity measures.
    \item \textbf{Speed:} Using the obtained embeddings, set similarities should be rapidly estimated, regardless of their cardinalities.
\end{itemize}

\begin{figure}[t]
	\centering
	\includegraphics[width=0.92\linewidth]{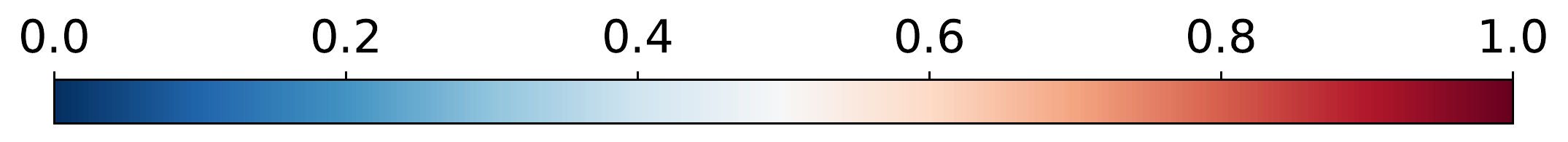}\\
	\begin{subfigure}[b]{.322\linewidth}
	    \centering
	    \captionsetup{justification=centering}
    	\includegraphics[width=0.99\linewidth]{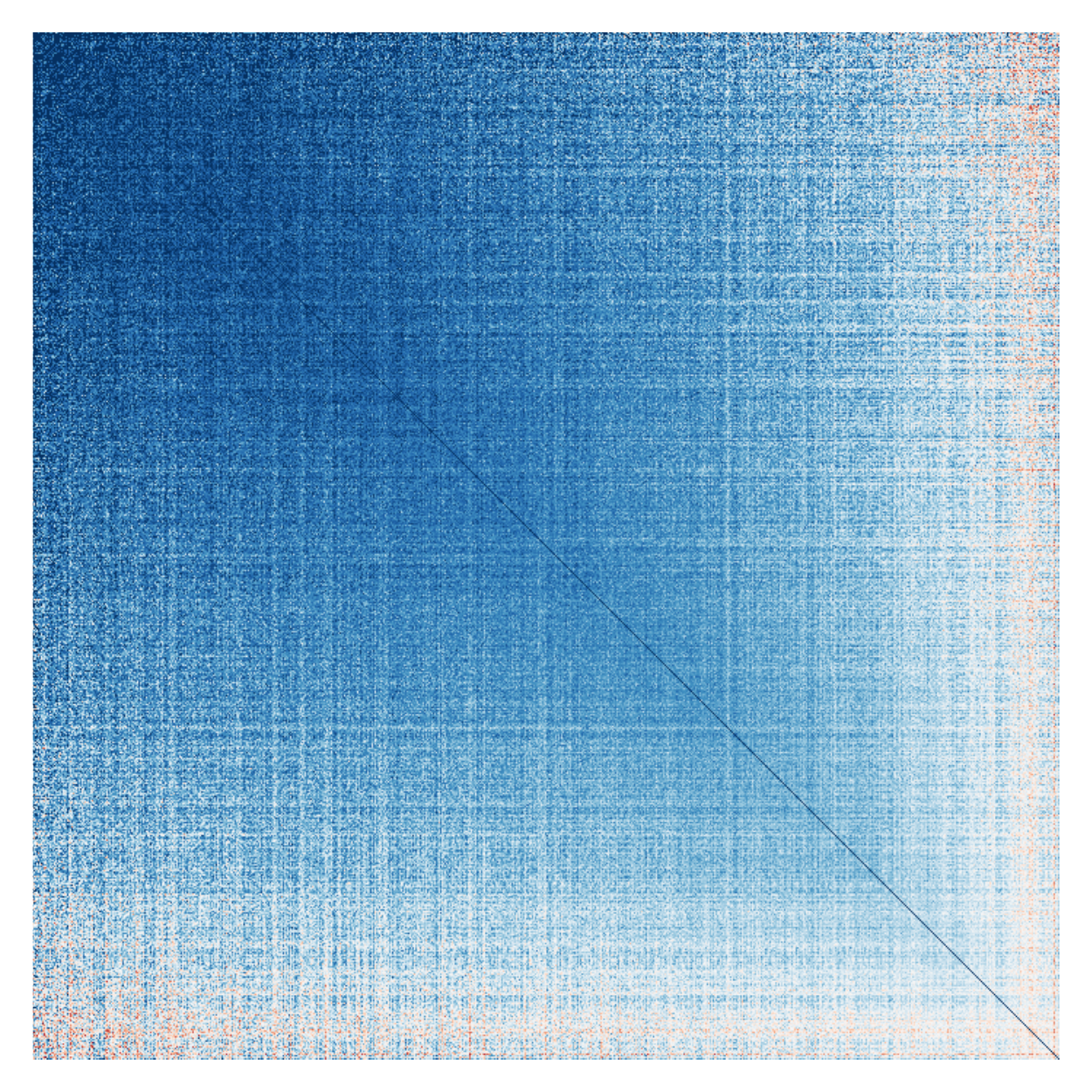}
    	\vspace{-15pt}
    	\caption{Random Hashing\\MSE = 0.0884\\Cost = 77.312 KB\label{fig:crown:random_hashing}}
	\end{subfigure}
	\hfill
	\begin{subfigure}[b]{.322\linewidth}
	    \centering
	    \captionsetup{justification=centering}
    	\includegraphics[width=0.99\linewidth]{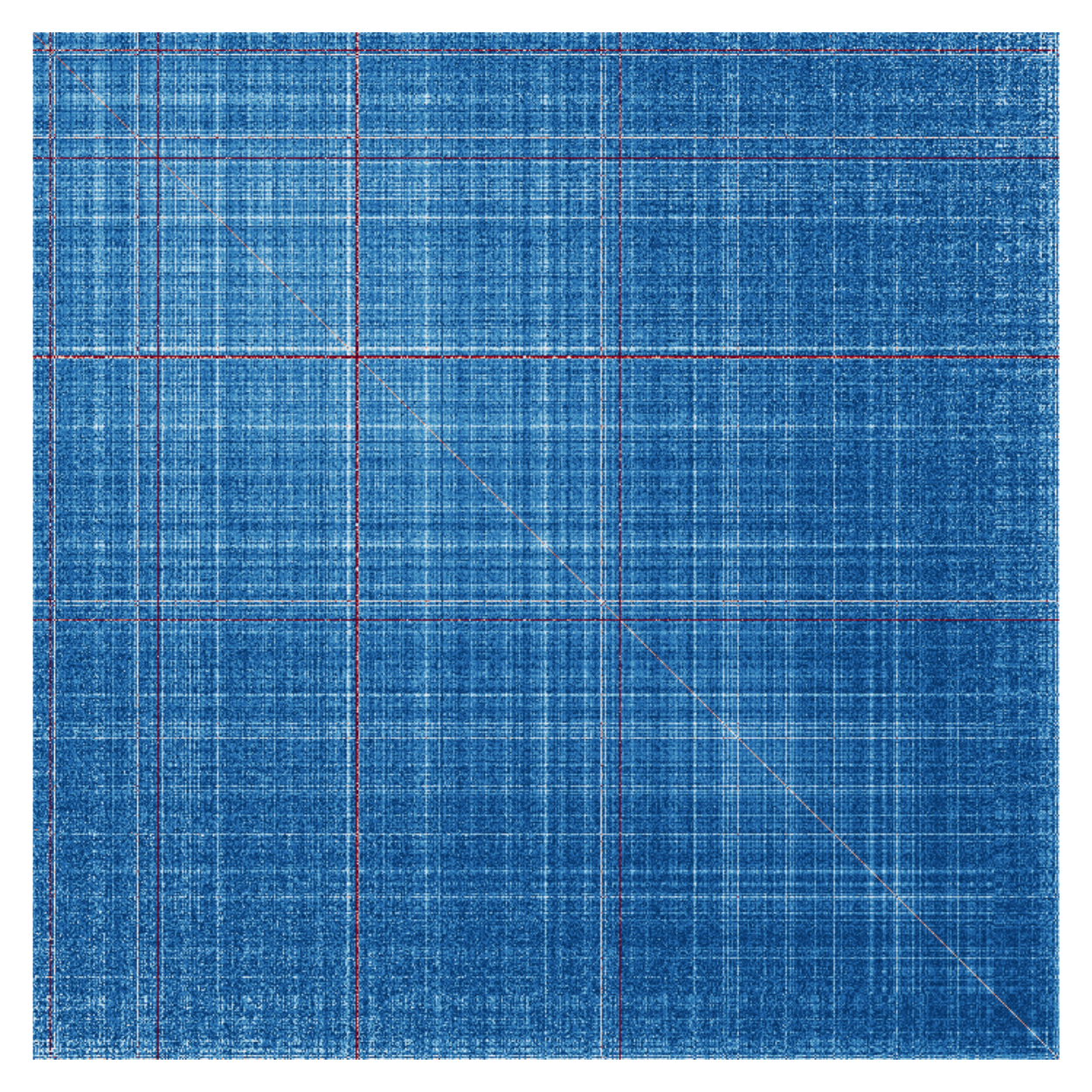}
    	\vspace{-15pt}
    	\caption{Vector Embedding\\MSE = 0.0495\\Cost = 77.312 KB}
	\end{subfigure}
	\hfill
	\begin{subfigure}[b]{.322\linewidth}
	    \centering
	    \captionsetup{justification=centering}
    	\includegraphics[width=0.99\linewidth]{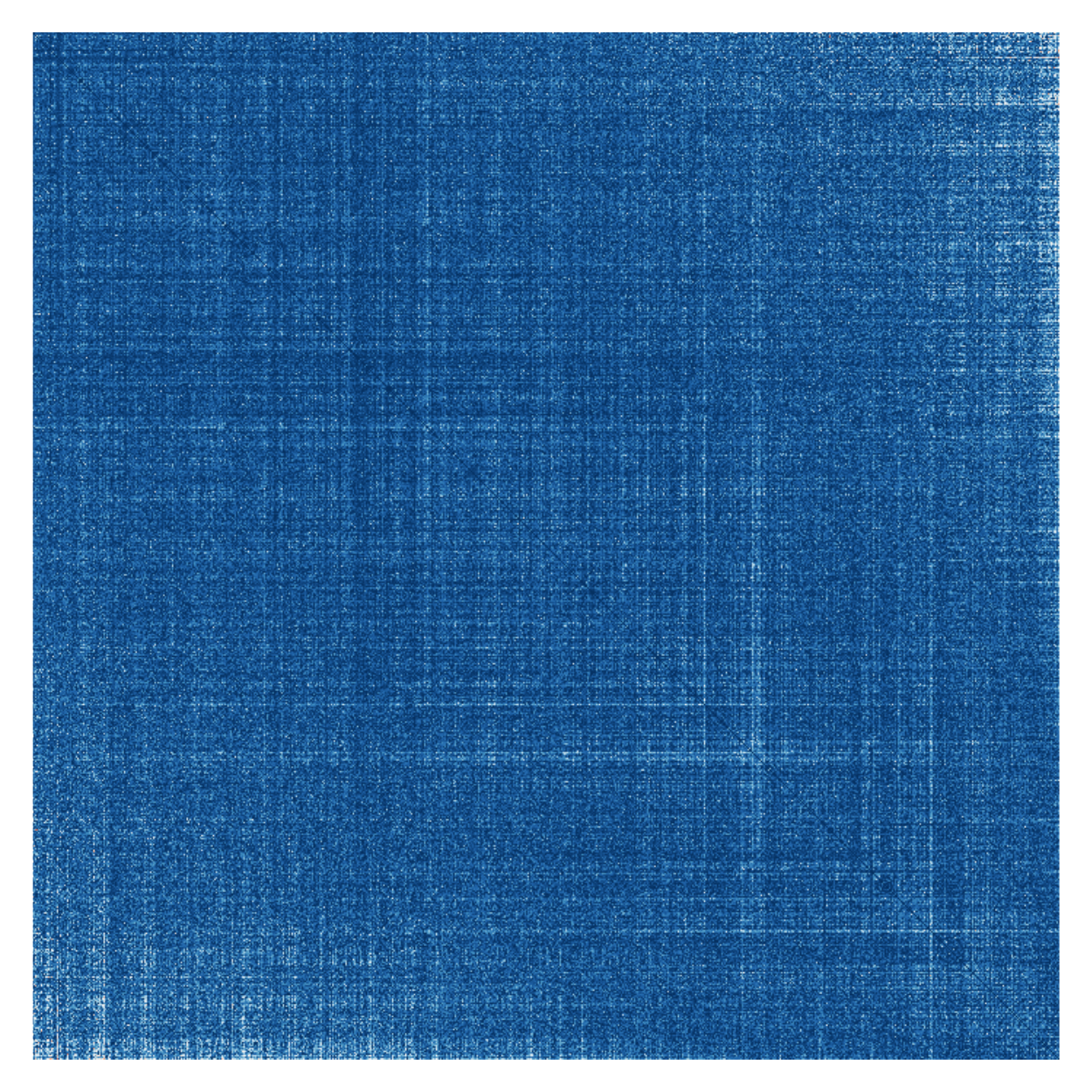}
    	\vspace{-15pt}
    	\caption{\methodq\\\textcolor{myred}{MSE = 0.0125}\\\textcolor{myred}{Cost = 15.695 KB}}
	\end{subfigure}
	\caption{Compared to intuitive methods, \methodq preserves the Overlap Coefficient between sets in the \textsf{MovieLens 1M} dataset more accurately while requiring smaller encoding cost. Rows and columns represent sets, and each cell represents the estimation error of pairwise set similarity. The indices of the sets are sorted by the sizes of the sets. \label{fig:crown}}
\end{figure}

\smallsection{Intuitive Methods:}
Keeping the above desirable properties in mind, we discuss simple and intuitive set-embedding methods for similarity preservation.

\begin{itemize}[leftmargin=*]
    \item \textbf{Random Hashing~\cite{guerraoui2020smaller}:}
    Each set $s$ is encoded as a binary vector $\mathrm{z}_s\in\{0,1\}^d$ by mapping each entity into one of the $d$ different values using a hash function $h(\cdot): \mathcal{E}\rightarrow \{1,\cdots,d\}$. Specifically, the representation $\mathrm{z}_s$ is derived by:
    \begin{equation*}
        \mathrm{z}_s[i] = 
        \begin{cases}
             1 & \text{if $\exists e\in s$ s.t. $h(e)=i$}\\
             0 & \text{otherwise.}
        \end{cases}   
    \end{equation*}
    The size of the set $s$ is estimated from the L1 norm (or the number of nonzero elements) of $\mathrm{z}_s$, i.e., $|s|\approx\|\mathrm{z}_s\|_1$. 
    In addition, sizes of the intersection and the union of sets $s$ and $s'$ are estimated from:
    \begin{equation*}
        |s\cap s'| \approx \|\mathrm{z}_s \;\textbf{AND}\; \mathrm{z}_{s'}\|_1 \;\;\;\text{and}\;\;\; |s\cup s'| \approx \|\mathrm{z}_s \;\textbf{OR}\; \mathrm{z}_{s'}\|_1,
    \end{equation*}
    respectively, where \textbf{AND} and \textbf{OR} are dimension-wise operations. 
    Based on these approximations, any set similarities (e.g., Jaccard Index) can be estimated.
    \item \textbf{Vector Embedding:}
    Another popular approach is to represent sets as vectors and compute the inner products between them to estimate a predefined set similarity.
    More precisely, given two sets $s$ and $s'$ and their vector representations $\mathrm{z}_s$ and $\mathrm{z}_{s'}$, it aims to approximate predefined $\text{sim}(s,s')$ by the inner product of $\mathrm{z}_s$ and $\mathrm{z}_{s'}$, i.e., $\langle \mathrm{z}_s, \mathrm{z}_{s'} \rangle\approx \text{sim}(s,s')$. 
\end{itemize}

These methods, however, suffer from several limitations. 
In random hashing, the maximum size of a set that a binary vector can accurately represent is $d$, and thus sets whose sizes are larger than $d$ inevitably suffer from information loss.
This is empirically verified in Figure~\ref{fig:crown:random_hashing}; while estimations are accurately made in small sets, the error increases as the sizes of the sets are larger.
The vector embedding method avoids such a problem but shows weakness in its versatility. 
That is, vectors are derived to preserve a predefined similarity (e.g., Jaccard Index), and thus they are not reusable to estimate other similarity measures (e.g., Dice Index). 
To address these issues, in this work, we propose \method and \methodq, novel end-to-end algorithms for similarity preserving set embedding. 
As shown in Figure~\ref{fig:crown}, \methodq accurately preserves similarities between sets compared to random hashing and vector embedding methods, while requiring fewer bits to encode sets.

	\section{Proposed Method}
	\label{sec:method}
	In this section, we present our proposed method for similarity-preserving set embedding.
We first present \method, a novel algorithm for learning similarity-preserving set representations using boxes (Sec.~\ref{sec:method:method}).
Then we propose \methodq, an advanced version of \method, which derives better conciseness and accuracy (Sec.~\ref{sec:method:methodq}).

\subsection{\method: Preliminary Version~\label{sec:method:method}}
How can we derive set embeddings that accurately preserve similarity in terms of various metrics?
Towards this goal, we first present \method, a preliminary set representation method that effectively learns the set itself and the structural relations with other sets.

\smallsection{Concepts:}
A \textit{box} is a $d$-dimensional hyper-rectangle whose representation consists of its center and offset~\cite{vilnis2018probabilistic}. 
The center describes the location of the box in the latent space and the offset is the length of each edge of the box. 
Formally, given a box $\mathrm{B}=(\mathrm{c}, \mathrm{f})$ whose center $\mathrm{c}\in \mathbb{R}^d$ and offset $\mathrm{f}\in \mathbb{R}_{+}^d$ are in the same latent space, the box is defined as a bounded region:
\begin{equation*}
    \mathrm{B} \equiv \{\mathrm{p}\in \mathbb{R}^{d} : \mathrm{c} - \mathrm{f} \preceq \mathrm{p} \preceq \mathrm{c} + \mathrm{f}\},
\end{equation*}
where $\mathrm{p}$ is any point within the box.
We let $\mathrm{m}\in \mathbb{R}^{d}$ and $\mathrm{M}\in\mathbb{R}^{d}$ be the vectors representing the minimum and the maximum at each dimension, respectively, i.e., $\mathrm{m}=\mathrm{c} - \mathrm{f}$ and $\mathrm{M}=\mathrm{c} + \mathrm{f}$.
Given two boxes $\mathrm{B}_{X}=(\mathrm{c}_X, \mathrm{f}_X)$ and $\mathrm{B}_{Y}=(\mathrm{c}_Y, \mathrm{f}_Y)$, the intersection is also a box, represented as:
\begin{align*}
    \mathrm{B}_X \cap \mathrm{B}_Y \!\! \equiv& \{\mathrm{p}\in \mathbb{R}^{d} : \mathbf{max}(\mathrm{m}_X, \mathrm{m}_Y) \preceq \mathrm{p} \preceq \mathbf{min}(\mathrm{M}_X, \mathrm{M}_Y)\}.
\end{align*}
The \textit{volume} $\mathbb{V}(\mathrm{B})$ of the box $\mathrm{B}$ is computed by the product of the length of an edge in each dimension, i.e., $\mathbb{V}(\mathrm{B}) = \prod_{i=1}^{d}(\mathrm{M}[i] - \mathrm{m}[i])$.
The volume of the union of the two boxes is simply computed by $\mathbb{V}(\mathrm{B}_{X}) + \mathbb{V}(\mathrm{B}_Y) - \mathbb{V}(\mathrm{B}_X\cap \mathrm{B}_Y)$.

\smallsection{Representation:}
The core idea of \method is to model each set $s$ as a box $\mathrm{B}_s = (\mathrm{c}_s, \mathrm{f}_s)$ so that the relations with other sets are properly preserved in the latent space.
To this end, \method approximates the volumes of the boxes to the relative sizes of the sets, i.e., $\mathbb{V}(\mathrm{B}_s) \propto |s|$.
In addition to the single-set level, \method aims to preserve the relations between different sets by approximating the volumes of the intersection of the boxes to the intersection sizes of the sets, i.e., $\mathbb{V}(\mathrm{B}_{s_i} \cap \mathrm{B}_{s_j}) \propto |s_i \cap s_j|$.
Notably, \method not only addresses limitations of random hashing and vector-based embeddings, but it also has various advantages benefited from unique properties of boxes, as we discuss in Section~\ref{sec:discussions}.

\smallsection{Objective:}
Now we turn our attention to how to capture such overlaps between sets using boxes. 
Recall that our goal is to derive \textit{accurate} and \textit{versatile} representations of sets, and towards the first goal, we take relations beyond pairwise into consideration. 
Specifically, we consider three different levels of set relations (i.e., single, pair, and triple-wise relations) to capture the underlying high-order structure of sets. 
In another aspect, we aim to derive versatile set representations that can be used to estimate various similarity measures (e.g., Jaccard Index and Dice Index).
With these goals in mind, we design an objective function that aims to preserve elemental relations among triple of sets.
Specifically, given a triple $\{s_i, s_j, s_k\}$ of sets, we consider seven cardinalities from three different levels of subsets:
\textbf{(1)} $|s_i|$, $|s_j|$, $|s_k|$, 
\textbf{(2)} $|s_i \cap s_j|$, $|s_j \cap s_k|$, $|s_k \cap s_i|$, and
\textbf{(3)} $|s_i \cap s_j \cap s_k|$ 
which contain single, pair, and triple-wise information, respectively, and we denote them from $c_1(s_i,s_j,s_k)$ to $c_7(s_i,s_j,s_k)$.
These seven elements fully-describe the relations among the three sets, and we argue that any similarity measures are computable using them.
In this regard, we aim to preserve the ratios of the seven cardinalities by the volumes of the boxes $\mathrm{B}_{s_i}$, $\mathrm{B}_{s_j}$, and $\mathrm{B}_{s_k}$ by minimizing the following objective:
\begin{multline*}
    \mathcal{J}(s_i,s_j,s_k, \mathrm{B}_{s_i}, \mathrm{B}_{s_j}, \mathrm{B}_{s_k}) = \\ \sum_{\ell=1}^{7}\left( p_\ell(s_i,s_j,s_k) - \hat{p}_\ell(\mathrm{B}_{s_i}, \mathrm{B}_{s_j}, \mathrm{B}_{s_k}) \right)^2,
\end{multline*}
where $p_\ell$ is the ratio of the $\ell$\textsuperscript{th} cardinality among the three sets (i.e., $p_\ell=c_\ell/\sum_{\ell'}c_{\ell'}$) and $\hat{p}_\ell$ is the corresponding ratio estimated by the boxes.
Since there exist $|\mathcal{S}| \choose 3$ possible triples of sets, taking all such combinations into account is practically intractable, and thus we resort to sampling some of them. 
We sample a set $\mathcal{T}$ of triples that consists of a set $\mathcal{T}^{+}$ of \textit{positive} triples and a set $\mathcal{T}^{-}$ of \textit{negative} triples, i.e., $\mathcal{T}=\mathcal{T}^{+}\cup \mathcal{T}^{-}$.
Specifically, the positive set $\mathcal{T}^{+}$ and the negative set $\mathcal{T}^{-}$ are obtained by sampling three connected (i.e., overlapping) sets and three uniform random sets, respectively.
Then, the final objective function we aim to minimize is:
\begin{equation}\label{eq:objective}
    \mathcal{L} = \sum\nolimits_{\{s_i, s_j, s_k\}\in \mathcal{T}}\mathcal{J}(s_i,s_j,s_k, \mathrm{B}_{s_i}, \mathrm{B}_{s_j}, \mathrm{B}_{s_k}).
\end{equation}
Notably, the proposed objective function aims to capture not only the pairwise interactions between sets, but also the triplewise relations to capture high-order overlapping patterns of the sets.
In addition, it does not rely on any predefined similarity measure, but is a general objective for learning key structural patterns of sets and their neighbors.
This prevents the model from overfitting to a specific measure and enables the model to yield accurate estimates to diverse metrics, as  shown empirically in Section~\ref{sec:experiments}.

\smallsection{Box Embedding:} 
Then, given a set $s$, how can we derive the box $\mathrm{B}_s = (\mathrm{c}_s, \mathrm{f}_s)$, that is, its center $\mathrm{c}_s$ and offset $\mathrm{f}_s$?
To make the method generalizable to unseen sets, \method introduces a pair of learnable embedding matrices $\mathbf{Q}^{\mathrm{c}}\in \mathbb{R}^{|\mathcal{E}|\times d}$ and $\mathbf{Q}^{\mathrm{f}}\in \mathbb{R}_{+}^{|\mathcal{E}|\times d}$ of entities, where $\mathbf{Q}_{i}^{\mathrm{c}}\in\mathbb{R}^d$ and $\mathbf{Q}_{i}^{\mathrm{f}}\in\mathbb{R}^d$ represent the center and offset of an entity $e_i$, respectively. 
Then, the embeddings of the entities in the set $s$ are 
aggregated to obtain the center $\mathrm{c}_s$ and the offset $\mathrm{f}_s$:
\begin{equation*}
    \mathrm{c}_s = \textbf{pooling}(s, \mathbf{Q}^{\mathrm{c}}) \;\;\;\text{and}\;\;\;
    \mathrm{f}_s = \textbf{pooling}(s, \mathbf{Q}^{\mathrm{f}})
\end{equation*}
where \textbf{pooling} is a permutation invariant function.
Instead of using simple functions such as mean or max, we use attentions to highlight the entities that are important to obtain either the center or the offset of the box.
To this end, we define a pooling function that takes the context of each set into account, termed set-context pooling (\textbf{SCP}).
Specifically, given a set $s$ and an item embedding matrix $\mathbf{Q}$ (which can be either $\mathbf{Q}^{\mathrm{c}}$ or $\mathbf{Q}^{\mathrm{f}}$), it first obtains the set-specific context vector $\mathrm{b}_s$:
\begin{equation*}
    \mathrm{b}_s = \sum_{e_i\in s}\alpha_{i}\mathbf{Q}_i \;\;\;\text{where}\;\;\; \alpha_{i}=\frac{\exp(\mathrm{a}^\intercal \mathbf{Q}_i)}{\sum_{e_j\in s}\exp(\mathrm{a}^\intercal \mathbf{Q}_j)}
\end{equation*}
where $\mathrm{a}$ is a global context vector shared by all sets. 
Then using the context vector $\mathrm{b}_s$, which specifically contains the information on set $s$, it obtains the output embedding from:
\begin{equation*}
    \textbf{SCP}(s, \mathbf{Q})=\sum_{e_i\in s}\omega_{i}\mathbf{Q}_i \;\;\text{where}\;\; \omega_{i}=\frac{\exp(\mathrm{b}_s^\intercal \mathbf{Q}_i)}{\sum_{e_j\in s}\exp(\mathrm{b}_s^\intercal \mathbf{Q}_j)}.
\end{equation*}
To be precise, $\mathrm{c}_s=\textbf{SCP}(s, \mathbf{Q}^{\mathrm{c}})$ and $\mathrm{f}_s=|s|^{\frac{1}{d}}\textbf{SCP}(s, \mathbf{Q}^{\mathrm{f}})$.
Note that, for the offset $\mathrm{f}_s$, we further take unique geometric properties of boxes into consideration.
For any entity $e_i\in s$, the subset relation $\{e_i\}\subseteq s$ holds, and thus the same condition $\mathrm{B}_{\{e_i\}}\subseteq \mathrm{B}_s$ for boxes is desired, which should satisfy $\mathrm{f}_{\{e_i\}} \preceq \mathrm{f}_s$ and thus $\max_{e_i\in s}\mathrm{f}_{\{e_i\}} = \max_{e_i\in s}\mathbf{Q}_i^{\mathrm{f}} \preceq \mathrm{f}_s$.
However, since SCP is the weighted mean of entities' embeddings, the output of SCP is bounded by the input embeddings in all dimensions, i.e., $\min_{e_i\in s}\mathbf{Q}_i^{\mathrm{f}} \preceq \mathrm{f}_s \preceq \max_{e_i\in s}\mathbf{Q}_i^{\mathrm{f}}$, which inevitably contradicts the aforementioned condition.
In these regards, for the offset $\mathrm{f}_s$, we multiply an additional regularizer $|s|^{\frac{1}{d}}$ that helps boxes to properly preserve the set similarity, i.e., $\mathrm{f}_s=|s|^{\frac{1}{d}}\textbf{SCP}(s, \mathbf{Q}^{\mathrm{f}})$.

\smallsection{Smoothing Boxes:}
By definition, a box $\mathrm{B}=(\mathrm{c}, \mathrm{f})$ is a bounded region with \textit{hard edges} whose volume is
\begin{equation*}
    \mathbb{V}(\mathrm{B})=\prod\nolimits_{i=1}^{d}(\mathrm{M}[i] - \mathrm{m}[i])=\prod\nolimits_{i=1}^{d}\text{ReLU}(\mathrm{M}[i] - \mathrm{m}[i])
\end{equation*}
where $\mathrm{m}=\mathrm{c}-\mathrm{f}$ and $\mathrm{M}=\mathrm{c}+\mathrm{f}$.
This, however, disables gradient-based optimization when boxes are disjoint~\cite{li2018smoothing}, and thus we \textit{smooth} the boxes by using an approximation of ReLU:
\begin{equation*}
    \mathbb{V}(\mathrm{B}) = \prod\nolimits_{i=1}^{d}\text{Softplus}(\mathrm{M}[i] - \mathrm{m}[i])
\end{equation*}
where $\text{Softplus}(\mathrm{x})=\frac{1}{\beta}\log \left(1 + \exp(\beta \mathrm{x})\right)$ is an approximation to $\text{ReLU}(\mathrm{x})$, and it becomes closer to ReLU as $\beta$ increases (specifically, Softplus $\rightarrow$ ReLU as $\beta\rightarrow \infty$).
In this way, any pairs of boxes overlap each other, and thus non-zero gradients are computed for optimization.

\smallsection{Encoding Cost:}
Each box consists of two vectors, a center and  an offset, and it requires $2\cdot 32d=64d$ bits to encode them, assuming that we are using float-32 to represent each real number. 
Thus, $64|\mathcal{S}|d$ bits are required to store the box embeddings of $|\mathcal{S}|$ sets.

\subsection{\methodq: Advanced Version~\label{sec:method:methodq}}
We describe \methodq, which enhances \method in terms of conciseness and accuracy, based on an end-to-end \textit{box quantization} scheme.
Specifically, \methodq compresses the the box embeddings into a compact set of key boxes and a set of discrete codes to reconstruct the original boxes.

\smallsection{Box Quantization:}
We propose box quantization, a novel scheme for compressing boxes by using substantially smaller number of bits. 
Note that conventional product quantization methods~\cite{chen2020differentiable}, which are for vector compression, are straightforwardly applicable, by independently reducing the center and the offset of the box.
However, it hardly makes use of geometric properties of boxes, and thus it does not properly reflect the complex relations between them.
The proposed box quantization scheme effectively addresses this issue through two steps: \textbf{(1)} box discretization and \textbf{(2)} box reconstruction.

\begin{figure}[t]
	\centering
	\begin{subfigure}[b]{.475\linewidth}
    	\includegraphics[width=0.99\linewidth]{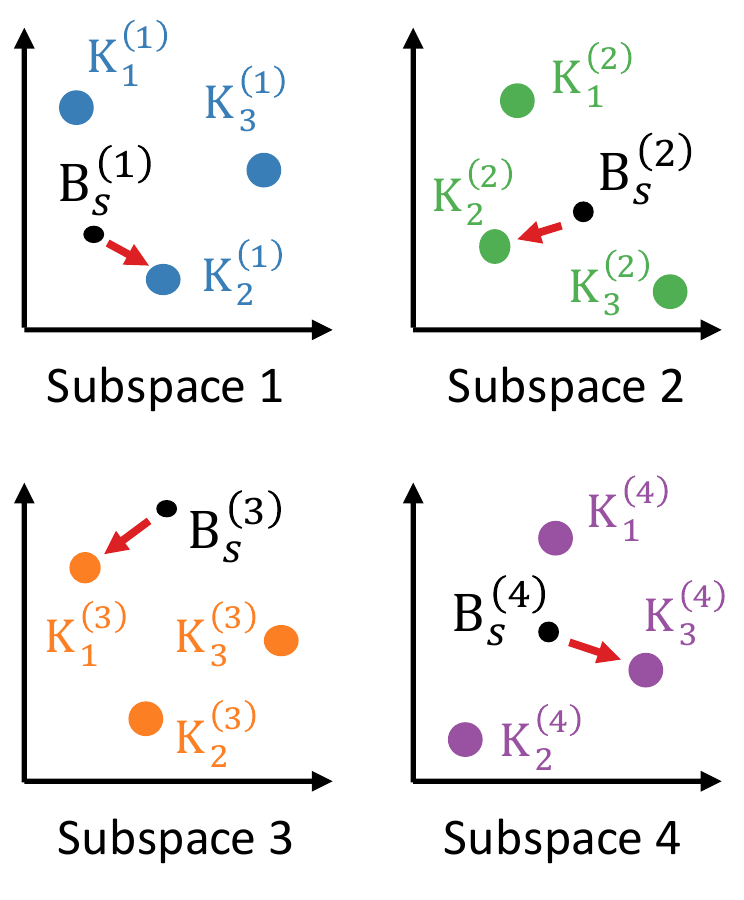}
    	\vspace{-15pt}
    	\caption{\small{Product Quantization}}
	\end{subfigure}
	\hfill
	\begin{subfigure}[b]{.475\linewidth}
    	\includegraphics[width=0.99\linewidth]{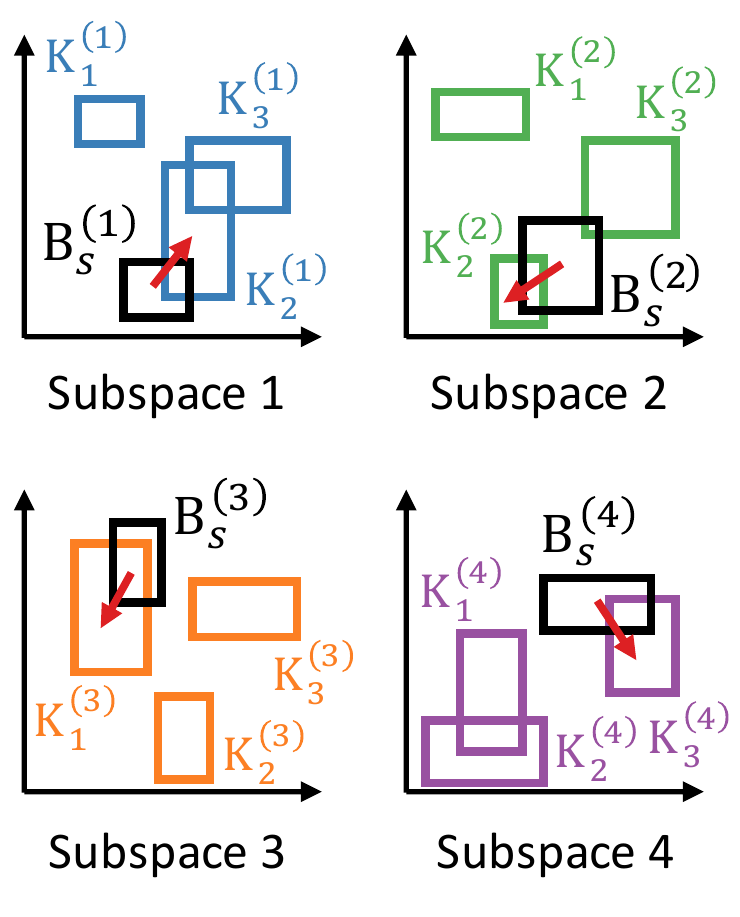}
    	\vspace{-15pt}
    	\caption{\small{Box Quantization}}
	\end{subfigure}
	\caption{An example of (a) product quantization and (b) box quantization when $D \;\text{(number of subspaces)} = 4$ and $K \;\text{(number of key boxes)} = 3$. While inner products of vectors are computed to measure the closeness in product quantization~\cite{chen2020differentiable}, the proposed box quantization scheme incorporates geometric relations between boxes. \label{fig:quantization}}
\end{figure}

\noindent
$\circ$ \textbf{Box Discretization.}
Given a box $\mathrm{B}_s=(\mathrm{c}_s, \mathrm{f}_s)$ of set $s$, we discretize the box as a $K$-way $D$-dimensional discrete code $\mathrm{C}_s\in \{1,\cdots, K\}^D$ which is more compact and requires much less number of bits to encode than real numbers.
To this end, we divide the $d$-dimensional latent space into $D$ subspaces ($\mathbb{R}^{d/D}$) and, for each subspace, learn $K$  \textit{key boxes}.
Specifically, in the $i$\textsuperscript{th} subspace, the $j$\textsuperscript{th} key box is denoted by $\mathrm{K}_{j}^{(i)}=(\mathrm{c}_j^{(i)}, \mathrm{f}_j^{(i)})$ where $\mathrm{c}_j^{(i)}\in \mathbb{R}^{d/D}$ and $\mathrm{f}_j^{(i)}\in \mathbb{R}_{+}^{d/D}$ are the center and offset of the key box, respectively.
The original box $\mathrm{B}_s$ is also partitioned into $D$ sub-boxes $\mathrm{B}_s^{(1)},\cdots,\mathrm{B}_s^{(D)}$ and the $i$\textsuperscript{th} code of $\mathrm{C}_s$ is decided by:
\begin{equation*}
    \mathrm{C}_{s}[i] = \argmin_j \textbf{dist}\left( \mathrm{B}_s^{(i)},\; \mathrm{K}_j^{(i)} \right)
\end{equation*}
where \textbf{dist}($\cdot,\cdot$) measures the distance (i.e., dissimilarity) between two boxes, and we can flexibly select the criterion. 
In this paper, we specify the \textbf{dist} function, using softmax, as:
\begin{equation}\label{eq:C_s}
    \mathrm{C}_{s}[i] = \argmax_j \frac{\exp\left(\textbf{BOR}\left(\mathrm{B}_s^{(i)}, \mathrm{K}_j^{(i)}\right)\right)}{\sum_{j'}\exp\left(\textbf{BOR}\left(\mathrm{B}_s^{(i)}, \mathrm{K}_{j'}^{(i)}\right)\right)}
\end{equation}
where \textbf{BOR} (Box Overlap Ratio) is defined to measure how much a box $\mathrm{B}_X$ and a box $\mathrm{B}_Y$ overlap:
\begin{equation*}
    \textbf{BOR}(\mathrm{B}_X, \mathrm{B}_Y) = \frac{1}{2}\left( \frac{\mathbb{V}(\mathrm{B}_X \cap \mathrm{B}_Y)}{\mathbb{V}(\mathrm{B}_X)} + \frac{\mathbb{V}(\mathrm{B}_X \cap \mathrm{B}_Y)}{\mathbb{V}(\mathrm{B}_Y)} \right).
\end{equation*}

As shown in Figure~\ref{fig:quantization}, the proposed box quantization scheme incorporates the geometric relations between boxes, differently from conventional product quantization methods on vectors.
To sum up, for each $i$\textsuperscript{th} subspace, we search for the key sub-box closest to the sub-box $\mathrm{B}_s^{(i)}$ and assign its index as the $i$\textsuperscript{th} dimension's value of its discrete code.

\noindent
$\circ$ \textbf{Box Reconstruction.}
Once the discrete code $\mathrm{C}_s$ of set $s$ is generated, in this step, we reconstruct the original box based on it.
To be specific, we obtain the reconstructed box $\widehat{\mathrm{B}}_s=(\widehat{c}_s, \widehat{f}_s)$ by concatenating $D$ key boxes from each subspace encoded in $\mathrm{C}_s$:
\begin{equation*}
    \widehat{\mathrm{B}}_s = \Big\Vert_{i=1}^{D} \mathrm{K}_{\mathrm{C}_s[i]}^{(i)}.
\end{equation*}
More precisely, $\widehat{\mathrm{B}}_s$ is reconstructed by concatenating the centers and the offsets of the $D$ key boxes respectively.
Since $\mathrm{C}_s$ encodes key boxes that largely overlap with the box $\mathrm{B}_s$ (i.e., high BOR), if properly encoded, we can expect the reconstructed box $\widehat{\mathrm{B}}_s$ to be geometrically similar to the original box $\mathrm{B}_s$.
In Section~\ref{sec:experiments:analysis}, we demonstrate the effectiveness of the proposed box quantization scheme by comparing it with the product quantization method.

\smallsection{Differentiable Optimization:}
Recall that \methodq is an end-to-end learnable algorithm, which requires all processes to be differentiable. 
However, the $\argmax$ operation in Eq.~\eqref{eq:C_s} is non-differentiable, and to this end, we utilize the softmax with the temperature $\tau$:
\begin{equation}\label{eq:C_diff}
    \widetilde{\mathrm{C}}_{s}[i] = \frac{\exp\left(\textbf{BOR}\left(\mathrm{B}_s^{(i)}, \mathrm{K}_j^{(i)}\right) / \tau\right)}{\sum_{j'}\exp\left(\textbf{BOR}\left(\mathrm{B}_s^{(i)}, \mathrm{K}_{j'}^{(i)}\right) / \tau\right)}.
\end{equation}
Note that $\widetilde{\mathrm{C}}_{s}[i]$ is a $K$-dimensional probabilistic vector whose $j$\textsuperscript{th} element indicates the probability for $\mathrm{K}_j^{(i)}$ being assigned as the closest key box, i.e., the probability of $\mathrm{C}_s[i]=j$.
Then, the key box $\widetilde{\mathrm{K}}_s^{(i)}=(\widetilde{\mathrm{c}}_s^{(i)}, \widetilde{\mathrm{f}}_s^{(i)})$ in the $i$\textsuperscript{th} subspace is the weighted sum of the $K$ key boxes:
\begin{equation*}
    \widetilde{\mathrm{K}}_s^{(i)} = \sum\nolimits_{j=1}^{K}\widetilde{C}_s[i][j] \cdot \mathrm{K}_{j}^{(i)}.
\end{equation*}
If $\tau = 0$, Eq.~\eqref{eq:C_diff} is equivalent to the $\argmax$ function, i.e., a one-hot vector where $\mathrm{C}_s[i]$\textsuperscript{th} dimension is 1 and others are 0.
In this case, $\widetilde{\mathrm{K}}_s^{(i)}$ becomes equivalent to $\mathrm{K}_{\mathrm{C}_{s}[i]}^{(i)}$, which is the exact reconstruction derivable from the discrete code $\mathrm{C}_s$.
However, since this \textit{hard selection} is non-differentiable and thus prevents an end-to-end optimization, we resort to the approximation by using the softmax with $\tau\neq 0$ which is fully differentiable. 
Specifically, we use different $\tau$s' in forward ($\tau=0$) and backward ($\tau =1$) passes, 
which effectively enables differentiable optimization.

\smallsection{Joint Training:}
For further improvement, we introduce a joint learning scheme in the box quantization scheme.
Given a triple $\{s_i,s_j,s_k\}$ of sets from the training data $\mathcal{T}$, we obtain their boxes $\mathrm{B}_{s_i}$, $\mathrm{B}_{s_j}$, and $\mathrm{B}_{s_k}$ and their reconstructed ones $\widehat{\mathrm{B}}_{s_i}$, $\widehat{\mathrm{B}}_{s_j}$, and $\widehat{\mathrm{B}}_{s_k}$ using the box quantization. 
While the basic version of \methodq optimizes the following objective:

\begin{table}[t]
    \begin{center}
    	\caption{\label{tab:datasets}Statistics of the 8 real-world datasets: the number of entities $|\mathcal{E}|$, the number of sets $|\mathcal{S}|$, the maximum set size $\textbf{max}_{s\in \mathcal{S}}|s|$, and the size of the dataset $\textbf{sum}_{s\in \mathcal{S}}|s|$.
    	}
    	\scalebox{0.935}{
    		\begin{tabular}{r|r|r|r|r}
    			\toprule
    			\textbf{Dataset} & $|\mathcal{E}|$ & $|\mathcal{S}|$ & $\textbf{max}_{s\in \mathcal{S}}|s|$ & $\textbf{sum}_{s\in \mathcal{S}}|s|$\\
    			\midrule
    			\textsf{Yelp} (\textsf{YP}) 
    			& 25,252 & 25,656 & 649 & 467K\\
    			\textsf{Amazon} (\textsf{AM}) 
    			& 55,700 & 105,655 & 555 & 858K\\
    			\textsf{Netflix} (\textsf{NF}) 
    			& 17,769 & 478,615 & 12,206 & 56.92M\\
    			\textsf{Gplus} (\textsf{GP}) 
    			& 107,596 & 72,271 & 5,056 & 13.71M\\
    			\textsf{Twitter} (\textsf{TW}) 
    			& 81,305 & 70,097 & 1,205 & 1.76M\\
    			\textsf{MovieLens 1M} (\textsf{ML1}) & 3,533 & 6,038 & 1,435 & 575K\\
    			\textsf{MovieLens 10M} (\textsf{ML10}) & 10,472 & 69,816 & 3,375 & 5.89M\\
    			\textsf{MovieLens 20M} (\textsf{ML20}) & 22,884 & 138,362 & 4,168 & 12.20M\\
    			\bottomrule %
    		\end{tabular}}
    	\end{center}
\end{table}

\begin{equation*}
    \mathcal{J}(s_i,s_j,s_j,\widehat{\mathrm{B}}_{s_i},\widehat{\mathrm{B}}_{s_j},\widehat{\mathrm{B}}_{s_k}),
\end{equation*}
we additionally make use of the original boxes during the optimization.
Specifically, we jointly train the original boxes together with the reconstructed ones so that both types of boxes can achieve high accuracy.
To this end, we consider the following eight losses:
\begin{align*}
    \mathcal{J}(s_i,s_j,s_j,{\mathrm{B}}_{s_i},{\mathrm{B}}_{s_j},{\mathrm{B}}_{s_k}), \;\;\;\;\;
    \mathcal{J}(s_i,s_j,s_j,\widehat{\mathrm{B}}_{s_i},{\mathrm{B}}_{s_j},{\mathrm{B}}_{s_k}),\\
    \mathcal{J}(s_i,s_j,s_j,{\mathrm{B}}_{s_i},\widehat{\mathrm{B}}_{s_j},{\mathrm{B}}_{s_k}), \;\;\;\;\;
    \mathcal{J}(s_i,s_j,s_j,{\mathrm{B}}_{s_i},{\mathrm{B}}_{s_j},\widehat{\mathrm{B}}_{s_k}),\\
     \mathcal{J}(s_i,s_j,s_j,\widehat{\mathrm{B}}_{s_i},\widehat{\mathrm{B}}_{s_j},{\mathrm{B}}_{s_k}), \;\;\;\;\;
    \mathcal{J}(s_i,s_j,s_j,\widehat{\mathrm{B}}_{s_i},{\mathrm{B}}_{s_j},\widehat{\mathrm{B}}_{s_k}),\\\mathcal{J}(s_i,s_j,s_j,{\mathrm{B}}_{s_i},\widehat{\mathrm{B}}_{s_j},\widehat{\mathrm{B}}_{s_k}), \;\;\;\;\;
    \mathcal{J}(s_i,s_j,s_j,\widehat{\mathrm{B}}_{s_i},\widehat{\mathrm{B}}_{s_j},\widehat{\mathrm{B}}_{s_k}),
\end{align*}
where we denote them by $\mathcal{J}_1$ to $\mathcal{J}_8$, for the sake of brevity. 
Notably, $\mathcal{J}_1$, which utilizes only the original boxes, is an objective used for \method, and $\mathcal{J}_8$ considers only the reconstructed boxes.
Based on these joint views from different types of boxes, the final loss function we aim to minimize is:
\begin{equation}\label{eq:methodq_loss}
    \mathcal{L} =\!\!\!\!\! \sum_{\{s_i,s_j,s_k\}\in \mathcal{T}}\!\!\!\!\! \lambda \left( \mathcal{J}_1 + \mathcal{J}_2 + \mathcal{J}_3 + \mathcal{J}_4 + \mathcal{J}_5 + \mathcal{J}_6 + \mathcal{J}_7 \right) + \mathcal{J}_8,
\end{equation}
where $\lambda$ is the coefficient for balancing the losses between the joint views and the loss from the reconstructed boxes.
In this way, both original boxes and the reconstructed ones are trained together to be properly located and shaped in the latent space.
Note that even though both types of boxes are jointly trained to achieve high accuracy, only the reconstructed boxes are used for inference.
We conduct ablation studies to verify the effectiveness of the joint training scheme in Section~\ref{sec:experiments:analysis}.

\smallsection{Encoding Cost:}
To encode the reconstructed boxes for each set, \methodq requires \textbf{(1)} key boxes and \textbf{(2)} discrete codes to encode each set. 
There exist $K$ key boxes in each of the $D$ subspaces whose dimensionality is $d/D$, which requires $64Kd$ bits to encode.
Each set is encoded as a $K$-way $D$-dimensional vector, which requires $D\log_2 K$ bits. 
To sum up, to encode $|\mathcal{S}|$ sets, \methodq requires $64Kd + |\mathcal{S}|D\log_2 K$ bits. 
Notably, if $K \ll |\mathcal{S}|$, then $64Kd$ bits are negligible, and typically, $D\log_2 K \ll 64d$ holds. 
Thus, the encoding cost of \methodq is considerably smaller than that of \method.

\smallsection{Similarity Computation:}
Once we obtain set representations, it is desirable to rapidly compute the estimated similarities in the latent space.
Boxes, which \method and \methodq derive, require constant time to compute a pairwise similarity between two sets, as formalized in Lemma~\ref{lma:computation}.
\begin{lma}[Time Complexity of Similarity Estimation]\label{lma:computation}
    Given a pair of sets $s$ and $s'$ and their boxes $\mathrm{B}_{s}$ and $\mathrm{B}_{s'}$, respectively, it takes $O(d)$ time to compute the estimated similarity $\widehat{\text{sim}}(\mathrm{B}_s,\mathrm{B}_{s'})$, where $d$ is a user-defined \textbf{constant} that does not depend on the sizes of $s$ and $s'$.
\end{lma}
\noindent\textbf{\textit{Proof.}}
Assume that the true similarity $\text{sim}(s,s')$ is computable using $|s|$, $|s'|$, and $|s\cap s'|$.
They are estimated by $\mathbb{V}(\mathrm{B}_{s})$, $\mathbb{V}(\mathrm{B}_{s'})$, and $\mathbb{V}(\mathrm{B}_{s}\cap \mathrm{B}_{s'})$, respectively, and each of them is computed by the product of $d$ values, which takes $O(d)$ time.
Hence, the total time complexity is $O(d)$. \hfill \qedsymbol

	\section{Experimental Results}
	\label{sec:experiments}
	\begin{figure*}[t]
	\centering
	 \includegraphics[width=0.525\textwidth]{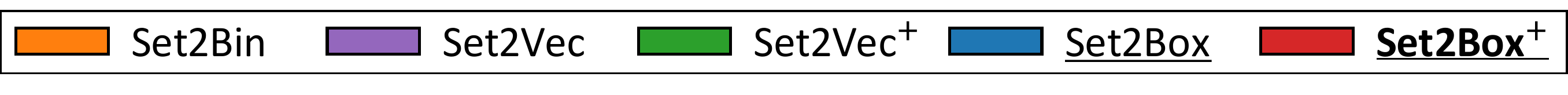}\\
	\begin{subfigure}[b]{.495\textwidth}
          \centering
          \captionsetup{justification=centering}
          \includegraphics[width=1.00\columnwidth]{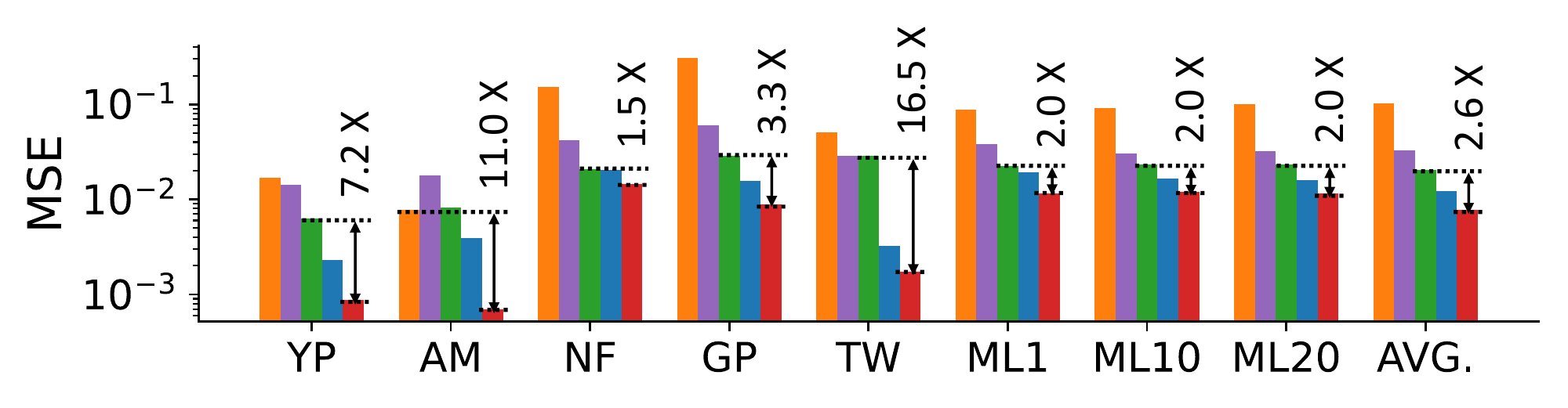}
		  \vspace{-16pt}
          \caption{Overlap Coefficient}
          \label{fig:result:oc}
    \end{subfigure}
    \hfill
    \begin{subfigure}[b]{.495\textwidth}
          \centering
          \captionsetup{justification=centering}
          \includegraphics[width=1.00\columnwidth]{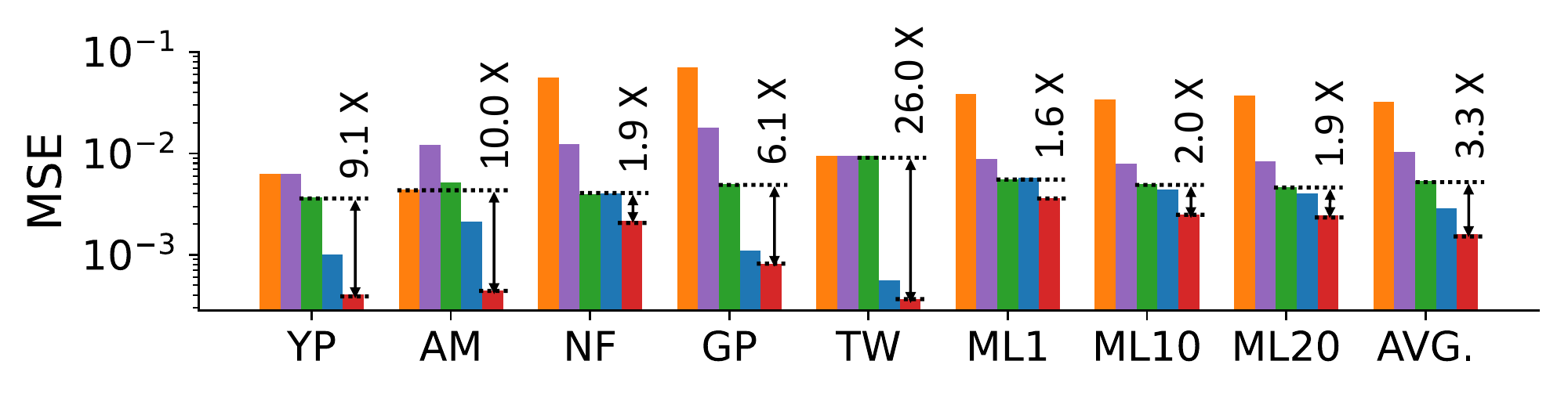}
		  \vspace{-16pt}
          \caption{Cosine Similarity}
          \label{fig:result:cs}
    \end{subfigure}
    \hfill
    \begin{subfigure}[b]{.495\textwidth}
          \centering
          \captionsetup{justification=centering}
          \includegraphics[width=1.00\columnwidth]{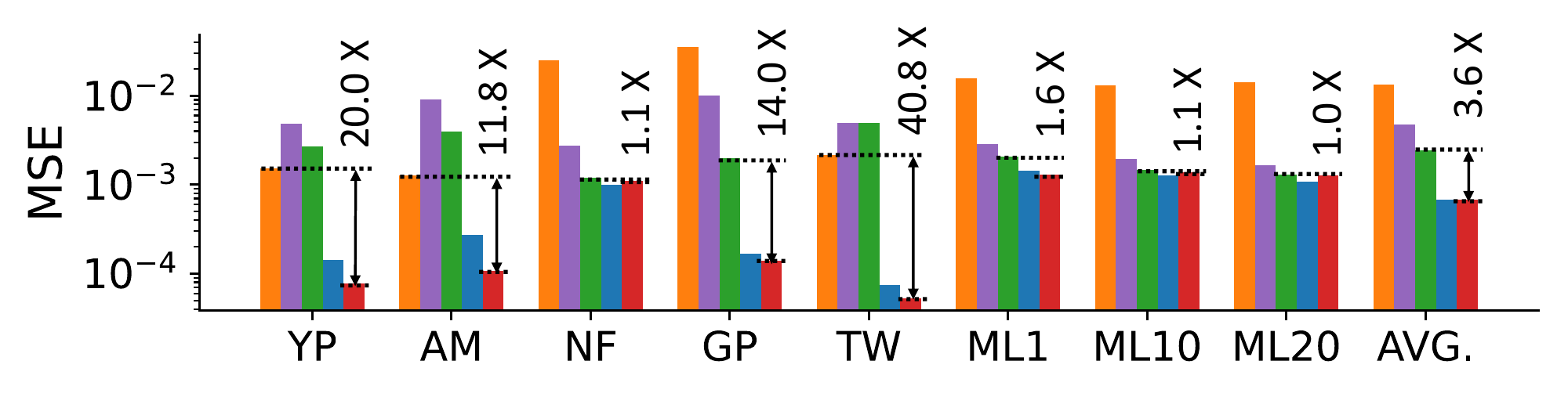}
		  \vspace{-16pt}
          \caption{Jaccard Index}
          \label{fig:result:ji}
    \end{subfigure}
    \hfill
    \begin{subfigure}[b]{.495\textwidth}
          \centering
          \captionsetup{justification=centering}
          \includegraphics[width=1.00\columnwidth]{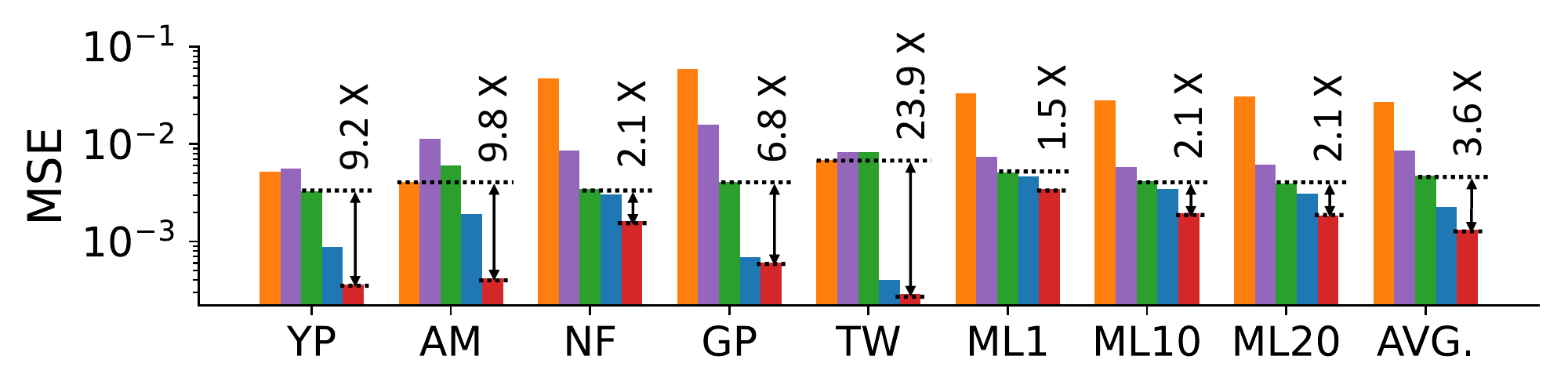}
		  \vspace{-16pt}
          \caption{Dice Index}
          \label{fig:result:di}
    \end{subfigure}
	\caption{\label{fig:result}\methodq preserves set similarities more accurately than \rh, \vect, \vectmlp, and \method. 
	Note that in \methodq, only $0.31 - 0.40$ of the bits costed by the competitors are used to embed sets. Moreover, while \vect and \vectmlp need to be  trained for specifically each similarity metric, \method and \methodq do not separate training.}
\end{figure*}

We review our experiments designed for answering Q1-Q3.

\begin{enumerate}[label=\textbf{Q\arabic*.},leftmargin=*]
    \item \textbf{Accuracy \& Conciseness:} Does \methodq derive concise and accurate set representations than its competitors? 
    \item \textbf{Effectiveness:} How does \methodq yield concise and accurate  representations? Are all its components useful?
    \item \textbf{Effects of Parameters:} How do the parameters of \methodq affect the quality of set representations?
\end{enumerate}

\subsection{Experimental Settings}
\smallsection{Machines \& Implementations:} 
All experiments were conducted on a Linux server with RTX 3090Ti GPUs.
We implemented all methods including \method and \methodq using the Pytorch library. 

\smallsection{Hyperparameter Tuning}
    Table~\ref{tab:hyperparameter} describes the hyperparameter search space of each method.
    The number of training samples, $|\mathcal{T}^{+}|$ and $|\mathcal{T}^{-}|$, are both set to $10$ for \method, \methodq, and their variants. 
    For the vector-based methods, \vect and \vectmlp, since three pairwise relations are extractable from each triple, $\lc\frac{7}{3} |\mathcal{T}^{+}|\rc$ positive triples and $\lc\frac{7}{3} |\mathcal{T}^{-}|\rc$ negative samples are used for training. 
    We fix the batch size to $512$ and use the Adam optimizer.
    In \methodq, we fix the softmax temperature $\tau$ to $1$.
    
\begin{table}[t]
    \begin{center}
    	\caption{\label{tab:hyperparameter}Search space of each method.}
    	\scalebox{0.975}{
    		\begin{tabular}{c|cc}
    			\toprule
    			\textbf{Method} & \textbf{Hyperparameter} & \textbf{Selection Pool}\\
    			\midrule
    			\multirow{2}{*}{\method} & Learning rate & $0.001$, $0.01$\\ 
    			& Box smoothing parameter $\beta$ & $1$, $2$, $4$\\
    			\midrule
    			\multirow{3}{*}{\methodq} & Learning rate & $0.001$, $0.01$\\ 
    			& Box smoothing parameter $\beta$ & $1$, $2$, $4$\\
    			& Joint training coefficient $\lambda$ & $0$, $0.001$, $0.01$, $0.1$, $1$\\
    			\bottomrule %
    		\end{tabular}}
    \end{center}
\end{table}

\begin{table}[t]
    	\begin{center}
    		\caption{\label{tab:cost}Encoding cost of the methods covered in this work. $|\mathcal{S}|$: number of sets. $d$: dimensionality. $D$: number of subspaces. $K$: number of key boxes (vectors) in each subspace.}
    		\scalebox{0.995}{
    			\begin{tabular}{c|ccc}
    				\toprule
    				\textbf{Method} && \textbf{Encoding Cost (bits)}&\\
    				\midrule
    				\rh && $d|\mathcal{S}|$ &\\
    				\vect && $32d|\mathcal{S}|$\\
    				\vectmlp && $32d|\mathcal{S}|$\\
    				\midrule
    				\methodo && $32d|\mathcal{S}|$ &\\
    				\methodpq && $64Kd + 2|\mathcal{S}|D\log_2 K$  &\\
    				\methodbq && $64Kd + |\mathcal{S}|D\log_2 K$  &\\
    				\midrule
    				\method && $64d|\mathcal{S}|$ &\\
    				\methodq && $64Kd + |\mathcal{S}|D\log_2 K$ &\\
    				\bottomrule %
    			\end{tabular}}
    	\end{center}
    \end{table}
    
\smallsection{Datasets:} 
We use eight publicly available datasets in Table~\ref{tab:datasets}.
The details of each dataset is as follows:
\begin{itemize}[leftmargin=*]
        \setlength\itemsep{-0.1em}
        \item \textbf{\textsf{\small{Yelp}} (\textsf{\small{YP}})} consists of user ratings on locations (e.g., hotels and restaurants), and each set is a group of locations that a user rated. Ratings higher than 3 are considered.
        \item \textbf{\textsf{\small{Amazon}} (\textsf{\small{AM}})} contains reviews of products (specifically, those categorized as Movies \& TV) by  users. In the dataset, each user has at least 5 reviews.
        A group of products reviewed by the same user is abstracted as a set.
        \item \textbf{\textsf{\small{Netflix}} (\textsf{\small{NF}})} is a collections of movie ratings from users. Each set is a set of movies rated by each user, and each entity is a movie. We consider ratings higher than 3.
        \item \textbf{\textsf{\small{Gplus}} (\textsf{\small{GP}})} is a directed social network that consists of `circles' from Google+. Each set is the group of neighboring nodes of each node.
        \item \textbf{\textsf{\small{Twitter}} (\textsf{\small{TW}})} is also a directed social networks consisting of `circles' in Twitter. Each set is a group of neighbors of each node in the graph.
        \item \textbf{\textsf{\small{MovieLens}} (\textsf{\small{ML1}}, \textsf{\small{ML10}}, and \textsf{\small{ML20}})} are collections of movie ratings from anonymous users. Each set is a group of movies that a user rated. Sets of movies with ratings higher than 3 are considered.
    \end{itemize}

\smallsection{Baselines:} 
We compare \method and \methodq with the following baselines including the variants of the methods discussed in Section~\ref{sec:prelim}: 
\begin{itemize}[leftmargin=*]
    \item \textbf{\rh} encodes each set $s$ as a binary vector $\mathrm{z}_s\in \{0,1\}^d$ using a random hash function. 
    See Section~\ref{sec:prelim} for details.
    \item \textbf{\vect} embeds each set $s$ as a vector $\mathrm{z}_s\in \mathbb{R}^d$ which is obtained by pooling learnable entity embeddings using SCP.
    Precisely, given two sets $s$ and $s'$ and their vector representations $\mathrm{z}_s$ and $\mathrm{z}_{s'}$, it aims to approximate the predefined set similarity $\text{sim}(s,s')$ by the inner product of $\mathrm{z}_s$ and $\mathrm{z}_{s'}$, i.e., $\langle \mathrm{z}_s, \mathrm{z}_{s'} \rangle\approx \text{sim}(s,s')$.
    \item \textbf{\vectmlp} incorporates entity features $\mathbf{X}\in\mathbb{R}^{|\mathcal{E}|\times d}$ into the set representation. 
    Features are projected using a fully-connected layer and then pooled to a set embedding.
\end{itemize}
In order to obtain the entity features for \vectmlp, which incorporates features into the set representations, we generate a projected graph (a.k.a., clique expansion) where nodes are entities, and any two nodes are adjacent if and only if their corresponding entities co-appear in at least one set.
Specifically, we generate a weighted graph by assigning a weight (specifically, the number of sets where the two corresponding entities co-appear) to each edge, and we apply node2vec~\cite{grover2016node2vec}, a popular random walk-based network embedding method, to the graph to obtain the feature of each entity.
Recall that vector-based methods, \vect and \vectmlp, are not versatile, and thus they need to be trained specifically to each metric, while the proposed methods \method and \methodq do not separate training. 
It should be noticed that search methods (e.g., LSH) are not direct competitors of the considered embedding methods, whose common goal is similarity preservation.
We summarize the encoding cost of each method, including the variants of \method and \methodq used in Section~\ref{sec:experiments:analysis}, in Table~\ref{tab:cost}.

\begin{figure*}[t]
    \centering
    \includegraphics[width=0.575\textwidth]{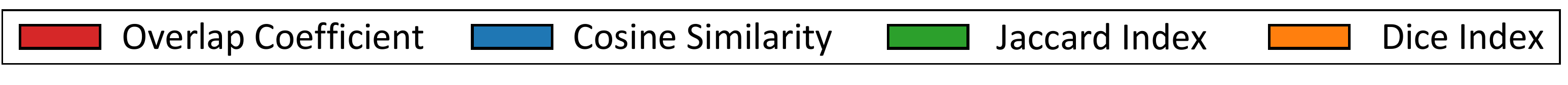}\\
    \begin{subfigure}[b]{.495\linewidth}
        \includegraphics[width=1.0\linewidth]{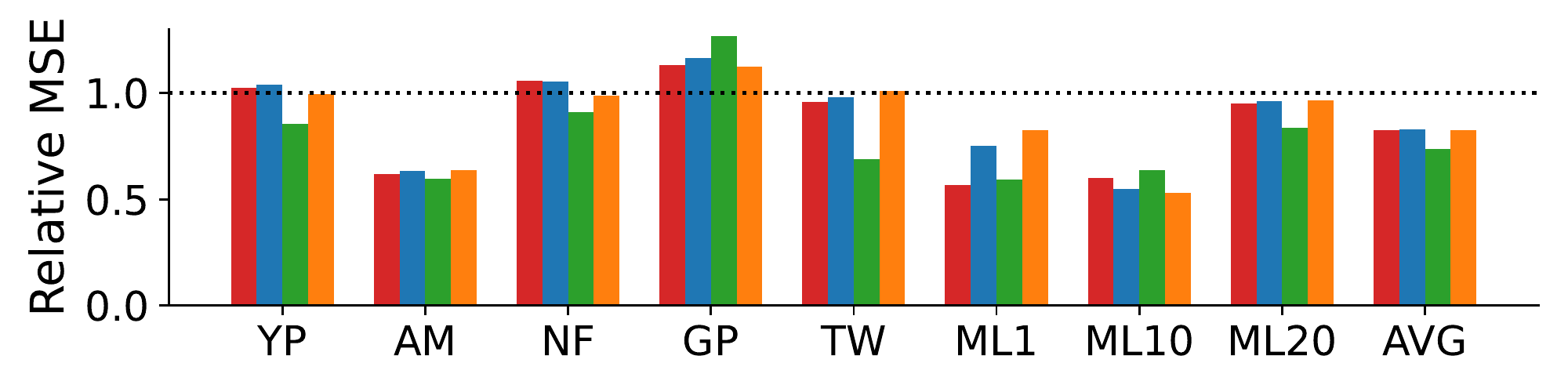}
        \vspace{-15pt}
        \caption{\small{Effects of the proposed box quantization scheme}\label{fig:effect:qbox}}
    \end{subfigure}
    \hfill
    \begin{subfigure}[b]{.495\linewidth}
        \includegraphics[width=1.0\linewidth]{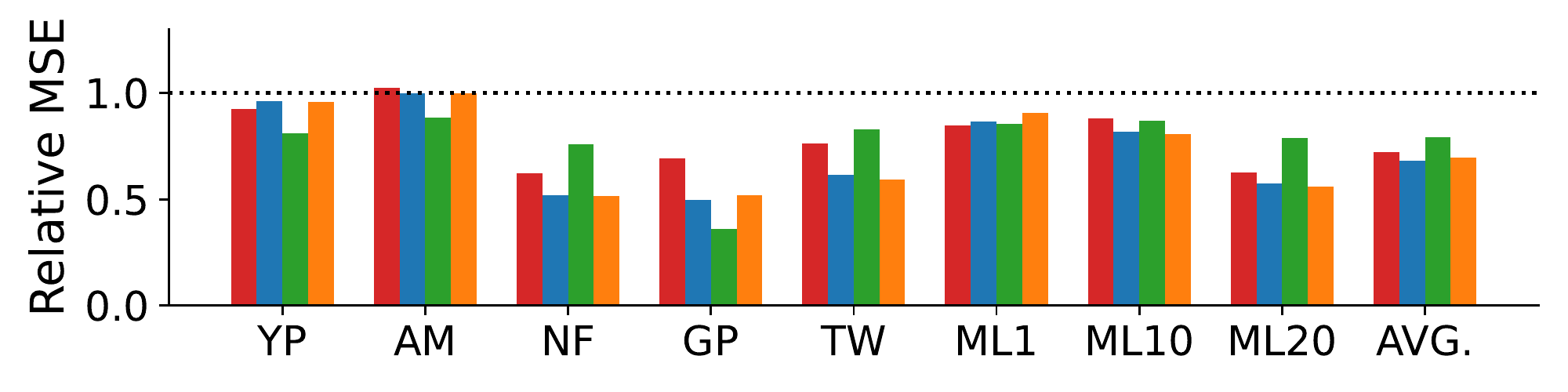}
        \vspace{-15pt}
        \caption{\small{Effects of joint training scheme}\label{fig:effect:joint}}
    \end{subfigure}
    \caption{Relative MSEs in Eq.~\eqref{eq:relmse:boxq} and \eqref{eq:relmse:joint} of the considered set similarity measures in each dataset. The proposed schemes: box quantization (\methodpq vs. \methodbq) and joint training (\methodbq vs. \methodq) improve the accuracy.\label{fig:effect}}
\end{figure*}

\smallsection{Evaluation:}  
For the \textsf{Netflix} dataset, whose number of sets is relatively large, we used $5\%$ of the sets for training.
For the other datasets, we used $20\%$ of the sets for training.
The remaining sets are split into half and used for validation and test.
We measured the Mean Squared Error (MSE) to evaluate the accuracy of the set similarity approximation.
Since the number of possible pairs of sets is $O(|\mathcal{S}|^2)$, which may be considerably large, we sample 100,000 pairs uniformly at random for evaluation.
We consider four representative set-similarity measures for evaluation: Overlap Coefficient (\textsc{OC}), Cosine Similarity (\textsc{CS}), Jaccard Index (\textsc{JI}), and Dice Index (\textsc{DI}), which are defined as:
\begin{equation*}
    \frac{|s \cap s'|}{\min(|s|, |s'|)}, \;\;\;  \frac{|s \cap s'|}{\sqrt{|s|\cdot |s'|}}, \;\;\; \frac{|s \cap s'|}{|s \cup s'|}, \;\;\;\text{and}\;\;\; \frac{|s \cap s'|}{\frac{1}{2}\left(|s| + |s'|\right)}, 
\end{equation*}
respectively, between a pair of sets $s$ and $s'$.

\subsection{Q1. Accuracy \& Conciseness\label{sec:experiments:accuracy_conciseness}}

We compare the MSE of the set similarity estimation derived by \methodq and its competitors. 
We set dimensions to 256 for \rh, 8 for vector based methods (\vect and \vectmlp), and 4 for \method, so that they use the same number of bits to encode sets.
For \methodq, we set $(d, D, K)=(32, 16, 30)$, which results in only $31-40\%$ of the encoding cost of the other methods, unless otherwise stated.

\smallsection{Accuracy:}
As seen in Figure~\ref{fig:result}, \methodq yields the most accurate set representations while using a smaller number of bits to encode them. 
For example, in the \textsf{Twitter} dataset, \methodq gives $40.8\times$ smaller MSE for the Jaccard Index compared to \rh. 
In the \textsf{Amazon} dataset, \methodq gives $11.0\times$ smaller MSE for the Overlap Coefficient than \vectmlp.
In both cases, \methodq uses about $31\%$ of the encoding costs used by the competitors.

\begin{figure}[t]
	\centering
	\includegraphics[width=0.4\textwidth]{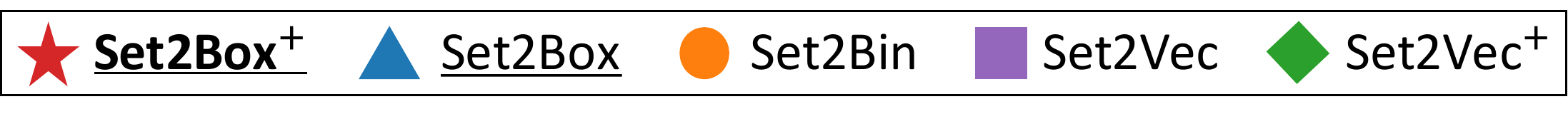}\\
	\begin{subfigure}[b]{.475\linewidth}
    	\includegraphics[width=1.0\linewidth]{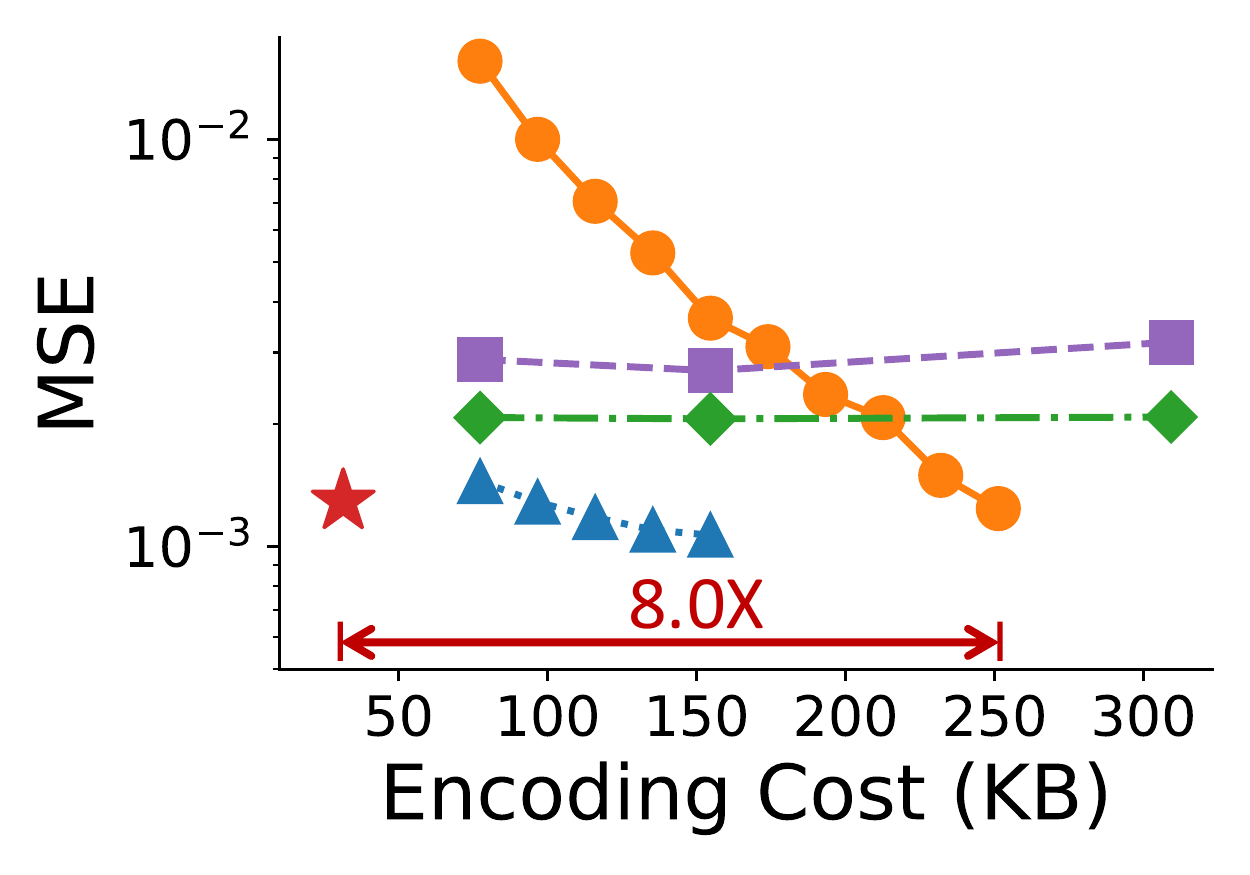}
    	\vspace{-15pt}
    	\caption{\textsf{MovieLens 1M}}
	\end{subfigure}
	\hfill
	\begin{subfigure}[b]{.475\linewidth}
    	\includegraphics[width=1.0\linewidth]{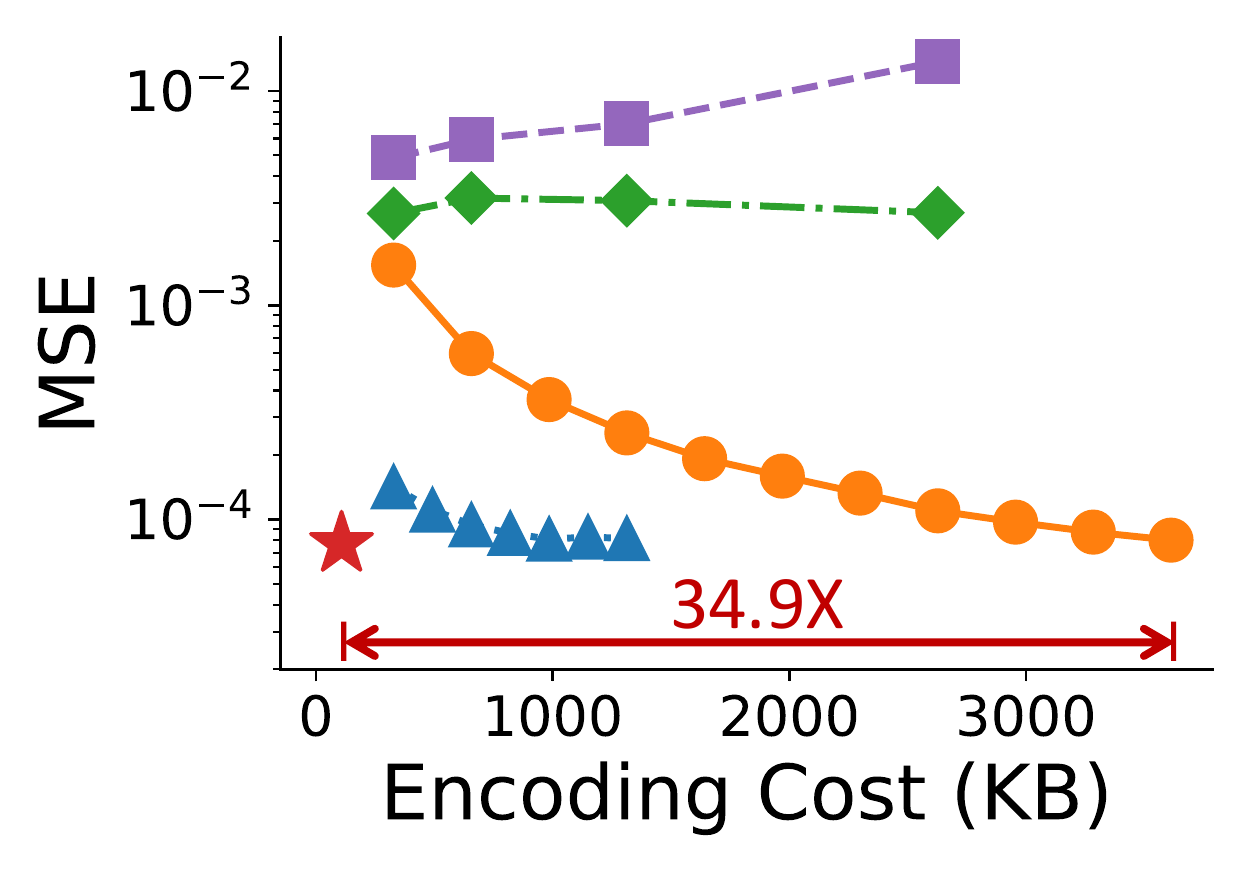}
    	\vspace{-15pt}
    	\caption{\textsf{Yelp}}
	\end{subfigure}\\
	\vspace{3pt}
	\begin{subtable}[h]{0.43\textwidth}
        \centering
        \begin{center}
		\scalebox{0.8}{
			\begin{tabular}{c|cccccccc}
				\toprule
				\textbf{Metric} & {\textsf{ML1}} & {\textsf{ML10}} & {\textsf{ML20}} & {\textsf{YP}} & {\textsf{AM}} & {\textsf{GP}} & {\textsf{TW}} & {\textsf{NF}}\\
				\midrule
				JI & 8.0 & 11.1 & 12.9 & 34.9 & 33.6 & 76.2 & 41.2 & 16.2\\
				DI & 8.0 & 15.9 & 17.7 & 27.3 & 27.2 & 63.5 & 31.7 & 22.7\\
				OC & 8.0 & 12.7 & 16.1 & 34.9 & 28.8 & 96.8 & 60.3 & 19.5\\
				CS & 8.0 & 15.9 & 16.1 & 28.8 & 28.8 & 68.2 & 38.0 & 22.7\\
				\bottomrule %
			\end{tabular}}
	    \end{center}
	    \vspace{-3mm}
        \caption{\label{tab:rh_dim}The number of times of the encoding cost that \rh requires to catch up the accuracy of \methodq.}
     \end{subtable}
	\caption{\methodq yields more concise set representations compared to the competitors.\label{fig:conciseness}}
\end{figure}

\begin{figure}[t]
	\centering
	\includegraphics[width=0.485\textwidth]{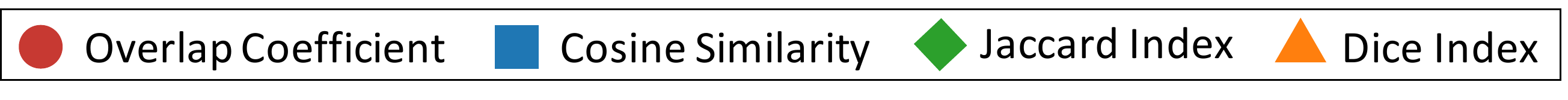}\\
	\begin{subfigure}[b]{.475\linewidth}
    	\includegraphics[width=1.0\linewidth]{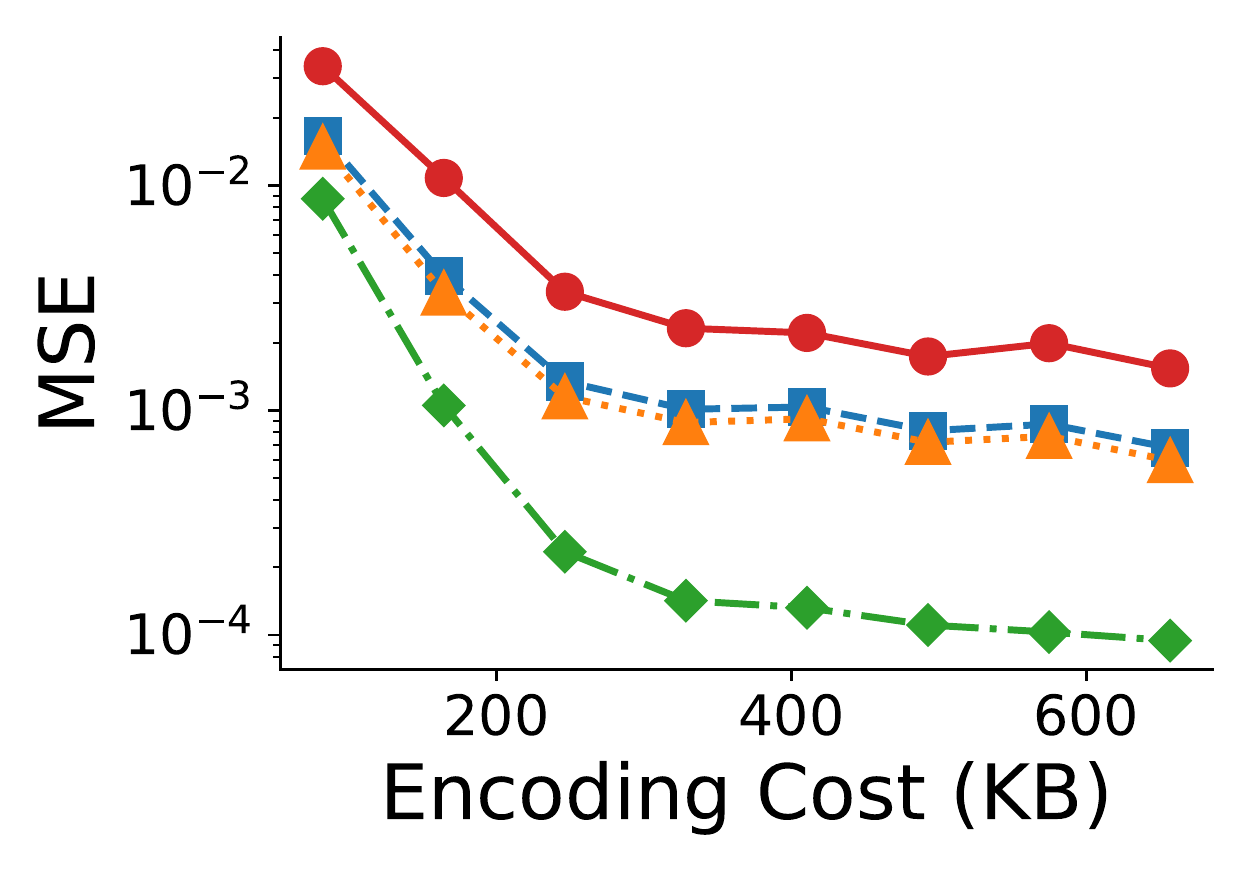}
    	\vspace{-15pt}
    	\caption{\method}
	\end{subfigure}
	\hfill
	\begin{subfigure}[b]{.475\linewidth}
    	\includegraphics[width=1.0\linewidth]{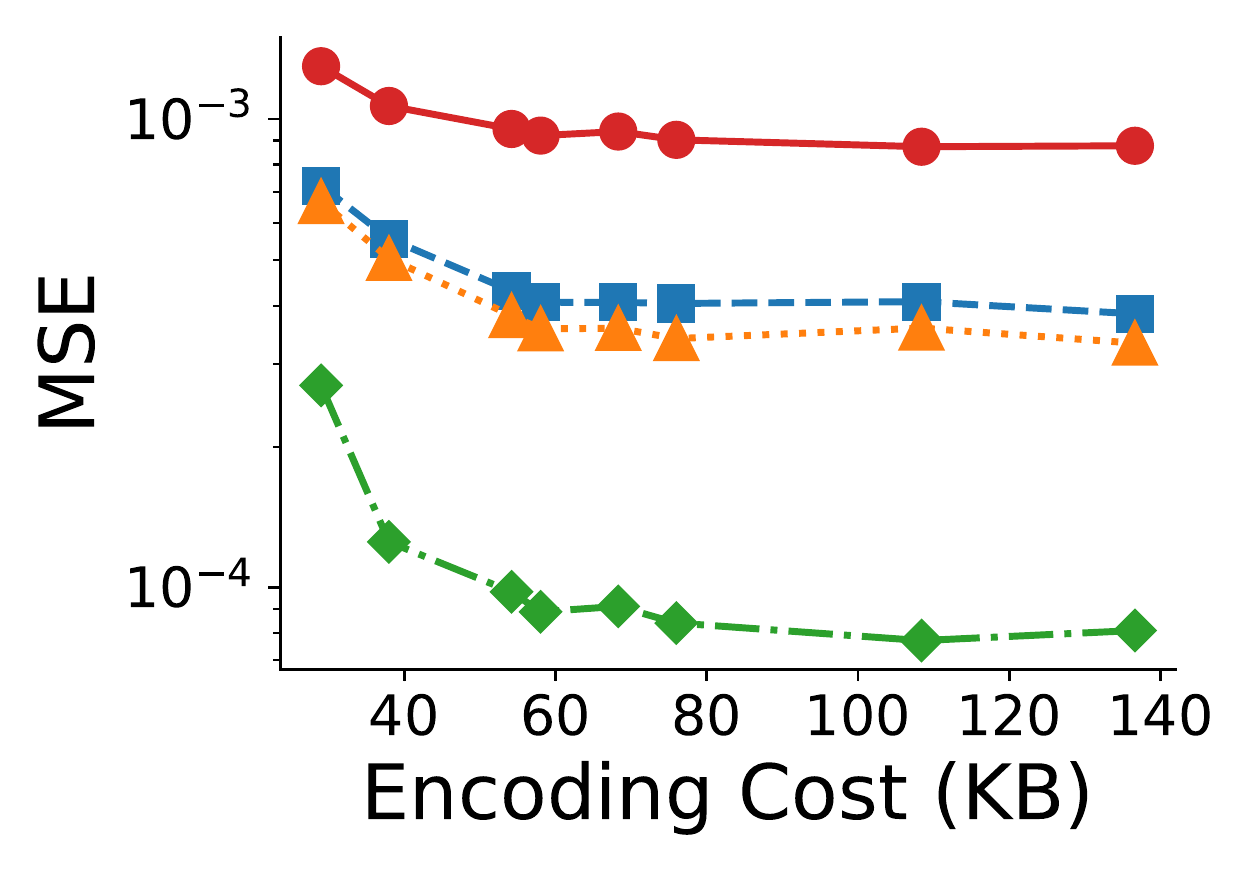}
    	\vspace{-15pt}
    	\caption{\methodq}
	\end{subfigure}
	\caption{\method and \methodq give more accurate estimations with larger space. \textsf{Yelp} is used for the plots.\label{fig:encoding_cost}}
\end{figure}

\smallsection{Conciseness:}
To verify the conciseness of \methodq, we measure the accuracy of competitors across various encoding costs.
As seen in Figure~\ref{fig:conciseness}, \methodq yields compact representations of sets while keeping them informative.
Vector-based methods are prone to the curse of dimensionality and hardly benefit from high dimensionality.
While the MSE of \rh decreases with respect to its dimension, it still requires larger space to achieve the MSE of \methodq.
For example, \rh requires $8.0\times$ and $34.9\times$ more bits to achieve the same accuracy of \methodq in \textsf{MovieLens 1M} and \textsf{Yelp}, respectively.
This is more noticeable in larger datasets, where \rh requires up to $96.8\times$ of the encoding cost of \methodq, as shown in Figure~\ref{tab:rh_dim}. These results demonstrate the conciseness of \methodq.
In addition, Figure~\ref{fig:encoding_cost} shows how the accuracy of estimations by \method and \methodq depend on their encoding costs.

\begin{figure}[t]
	\centering
	\includegraphics[width=0.38\textwidth]{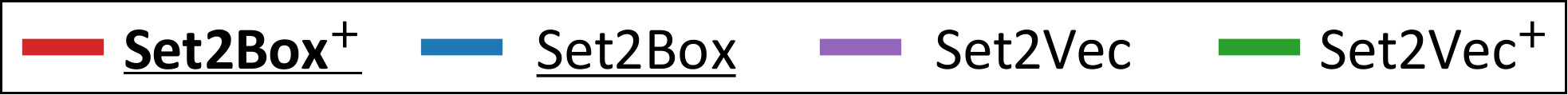}\\
	\begin{subfigure}[b]{.475\linewidth}
    	\includegraphics[width=1.0\linewidth]{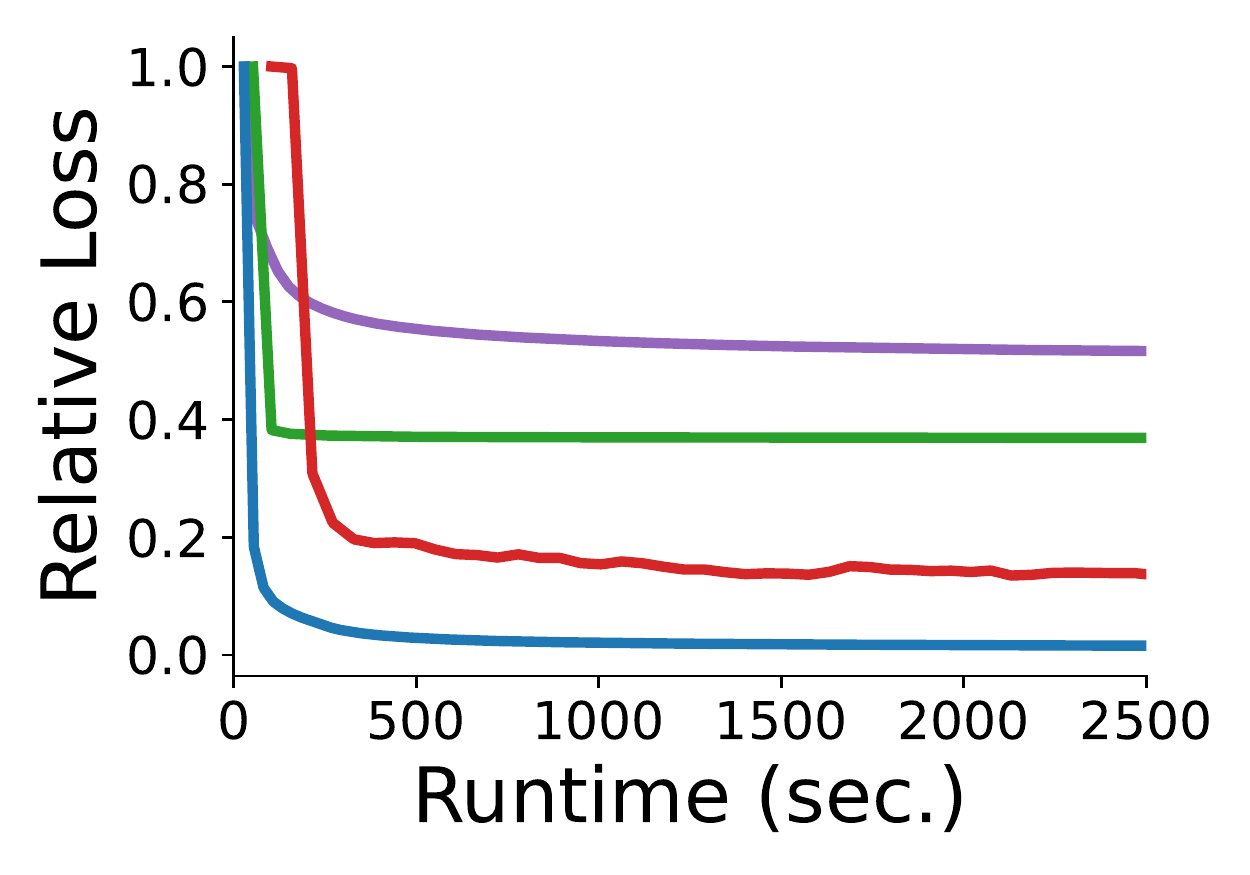}
    	\vspace{-15pt}
    	\caption{\textsf{MovieLens 20M}}
	\end{subfigure}
	\hfill
	\begin{subfigure}[b]{.475\linewidth}
    	\includegraphics[width=1.0\linewidth]{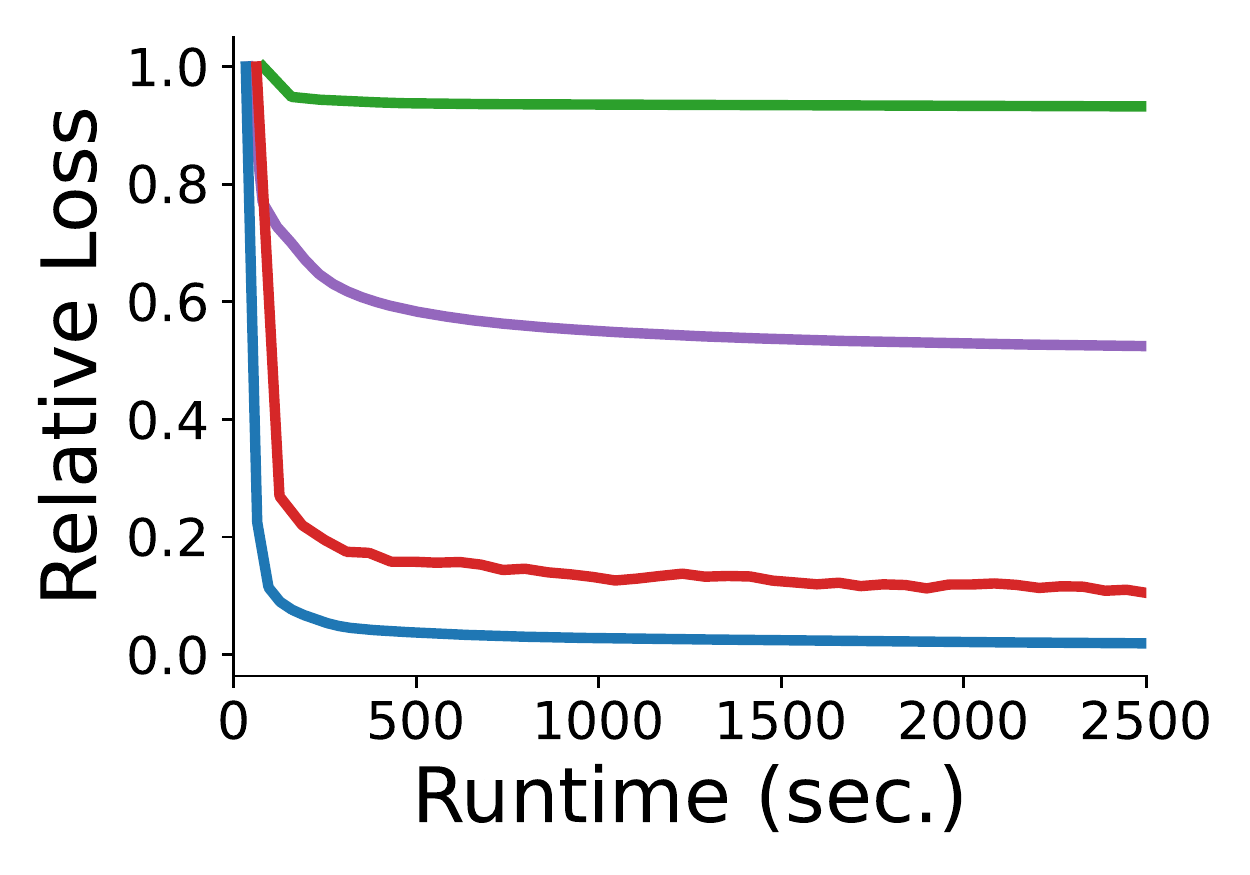}
    	\vspace{-15pt}
    	\caption{\textsf{Netflix}}
	\end{subfigure}
	\caption{\methodq converges over time.\label{fig:runtime}}
\end{figure}

\smallsection{Speed:}
For the considered methods,
Figure~\ref{fig:runtime} shows the loss (relative to the loss after the first epoch) over time in two large datasets, \textsf{MovieLens 20M} and \textsf{Netflix}.
The loss in \methodq drops over time and eventually converges 
within an hour.

\begin{figure}[t]
	\centering
	\includegraphics[width=0.42\textwidth]{FIG/legend/plot_legend.pdf}\\
	\begin{subfigure}[b]{.475\linewidth}
    	\includegraphics[width=1.0\linewidth]{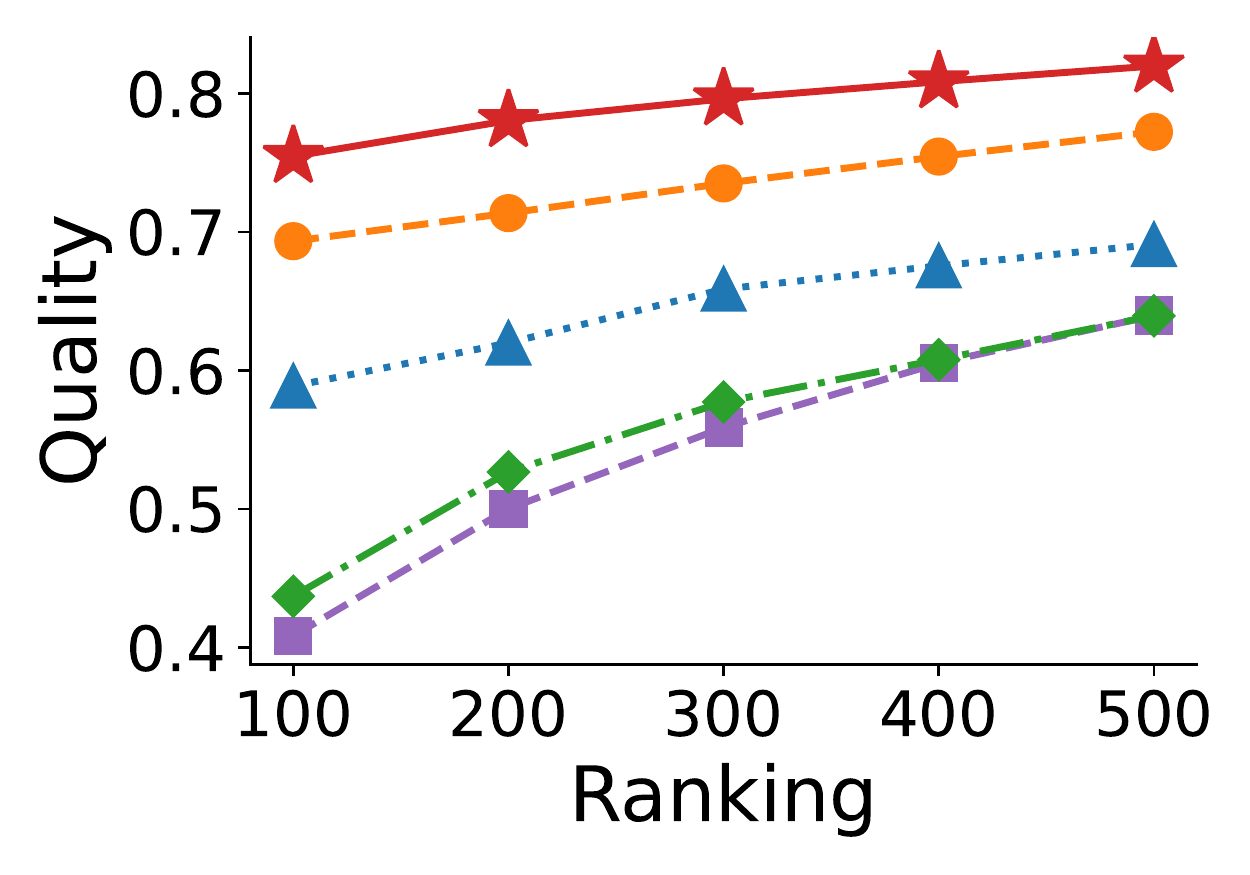}
    	\vspace{-15pt}
    	\caption{\textsf{MovieLens 1M}}
	\end{subfigure}
	\hfill
	\begin{subfigure}[b]{.475\linewidth}
    	\includegraphics[width=1.0\linewidth]{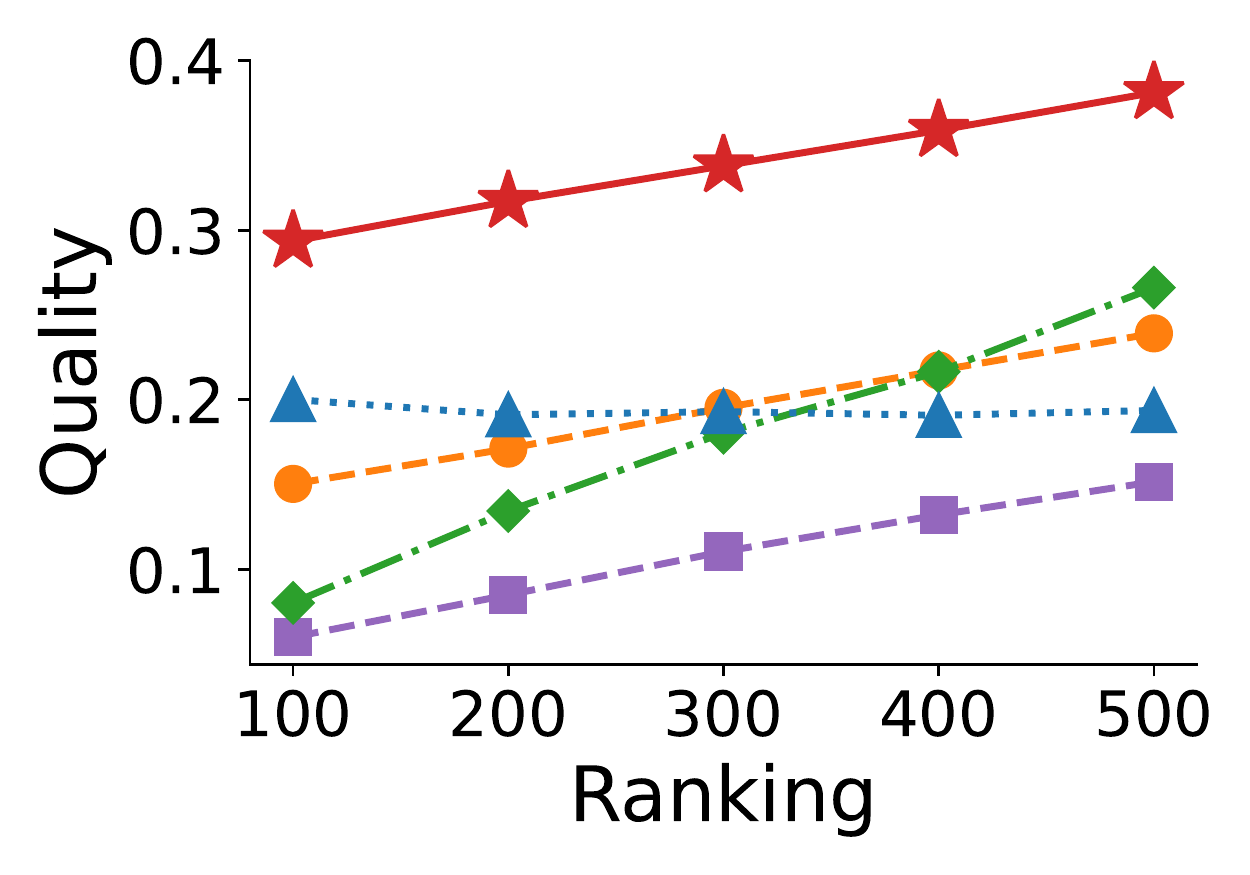}
    	\vspace{-15pt}
    	\caption{\textsf{Yelp}}
	\end{subfigure}
	\caption{
	Sets estimated to be similar by \methodq are indeed similar to each other\label{fig:knn}.}
\end{figure}

\smallsection{Further Analysis of Accuracy:}
We analyze how much sets estimated to be similar by the considered methods are actually similar.
To this end, for each set $s$, we compare its $k$ most similar sets $G_{s,k}$ to those $\widehat{G}_{s,k}$ estimated to be the most similar using each method.
Then, as in \cite{guerraoui2020smaller}, we measure the quality $q(\widehat{G}_{s,k})$ of $\widehat{G}_{s,k}$ which measures how sets in $\widehat{G}_{s,k}$ are similar enough to $s$ compared to those in $G_{s,k}$:
\begin{equation*}
    q(\widehat{G}_{s,k}) = \frac{\sum_{s'\in \widehat{G}_{s,k}}\text{sim}(s,s')}{\sum_{s'\in G_{s,k}}\text{sim}(s,s')}
\end{equation*}
where $\text{sim}(\cdot,\cdot)$ is the similarity (spec., Jaccard Index) between sets.
The quality ranges from $0$ to $1$, and it is ideally close to 1, which indicates that the sets estimated to be similar by the method are similar enough compared to the ideal sets.
Based on the above criterion, we measure the quality of each method with different $k$'s.
We use the above configuration for \methodq, while the dimensions of the other methods are adjusted to require a similar encoding cost to \methodq.
As shown in Figure~\ref{fig:knn}, the average quality $q(\widehat{G}_{s,k})$ is the highest in \methodq in all considered $k$'s in \textsf{MovieLens 1M} and \textsf{Yelp}, implying that the sets estimated to be similar by the proposed methods are indeed similar to each other.

\subsection{Q2. Effectiveness\label{sec:experiments:analysis}}
To verify the effectiveness of each component of \methodq, we conduct ablation studies by comparing it with its variants.
We first consider the following variants:
\begin{itemize}[leftmargin=*]
    \item \textbf{\methodpq:} Given a box $\mathrm{B}=(\mathrm{c},\mathrm{f})$, we apply an end-to-end differentiable product quantization (PQ)~\cite{chen2020differentiable} to the center $\mathrm{c}$ and the offset $\mathrm{f}$ independently. Dot products between the query vector ($\mathrm{c}$ or $\mathrm{r}$) and the key vectors are computed to measure the distances.
    Notably, it yields two independent discrete codes for the center and the offset, and thus its encoding cost is $64Kd+2|\mathcal{S}|D \log_2 K$ bits, which is approximately twice that of \methodq.
    \item \textbf{\methodbq:} A special case of \methodq with $\lambda=0$, where the proposed box quantization is applied but joint training is not.
\end{itemize}
We set $(d,D,K)$ to $(32,8,30)$ for \methodpq and $(32,16,30)$ for \methodbq and \methodq so they all require the the same amount of storage. 

\smallsection{Effects of Box Quantization:}
We examine the effectiveness of the proposed box quantization scheme in Section~\ref{sec:method:methodq} by comparing \methodbq with \methodpq. 
To this end, we measure the relative MSE defined as:
\begin{equation}\label{eq:relmse:boxq}
    \frac{\text{MSE of \methodbq}}{\text{MSE of \methodpq}}
\end{equation}
of each dataset.
Figure~\ref{fig:effect:qbox} demonstrates that \methodbq generally derives more accurate set representations compared to \methodpq, implying the effectiveness of the proposed box quantization scheme.
As shown in Table~\ref{tab:effects:methodq}, on average, \methodbq yields up to $26\%$ smaller MSE than \methodpq while using approximately the same number of bits.
For example, in \textsf{MovieLens 10M}, \methodbq gives $1.89\times$ more accurate estimation than \methodpq in approximating the Dice Index.
While \methodpq discretizes the center and offset of the boxes independently, without the consideration of their geometric properties, the proposed box quantization scheme effectively takes the geometric relations between boxes into account and thus yields high-quality compression.

\begin{figure}[t]
	\centering
	\includegraphics[width=0.265\textwidth]{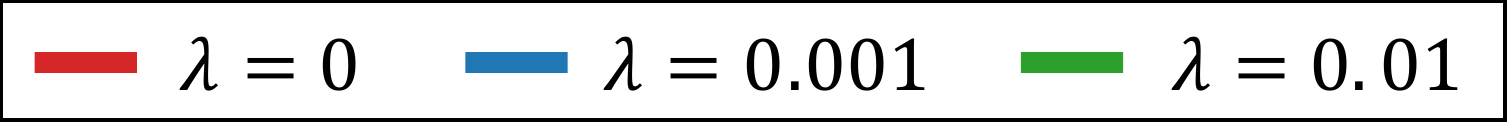}\\
	\begin{subfigure}[b]{.475\linewidth}
    	\includegraphics[width=1.0\linewidth]{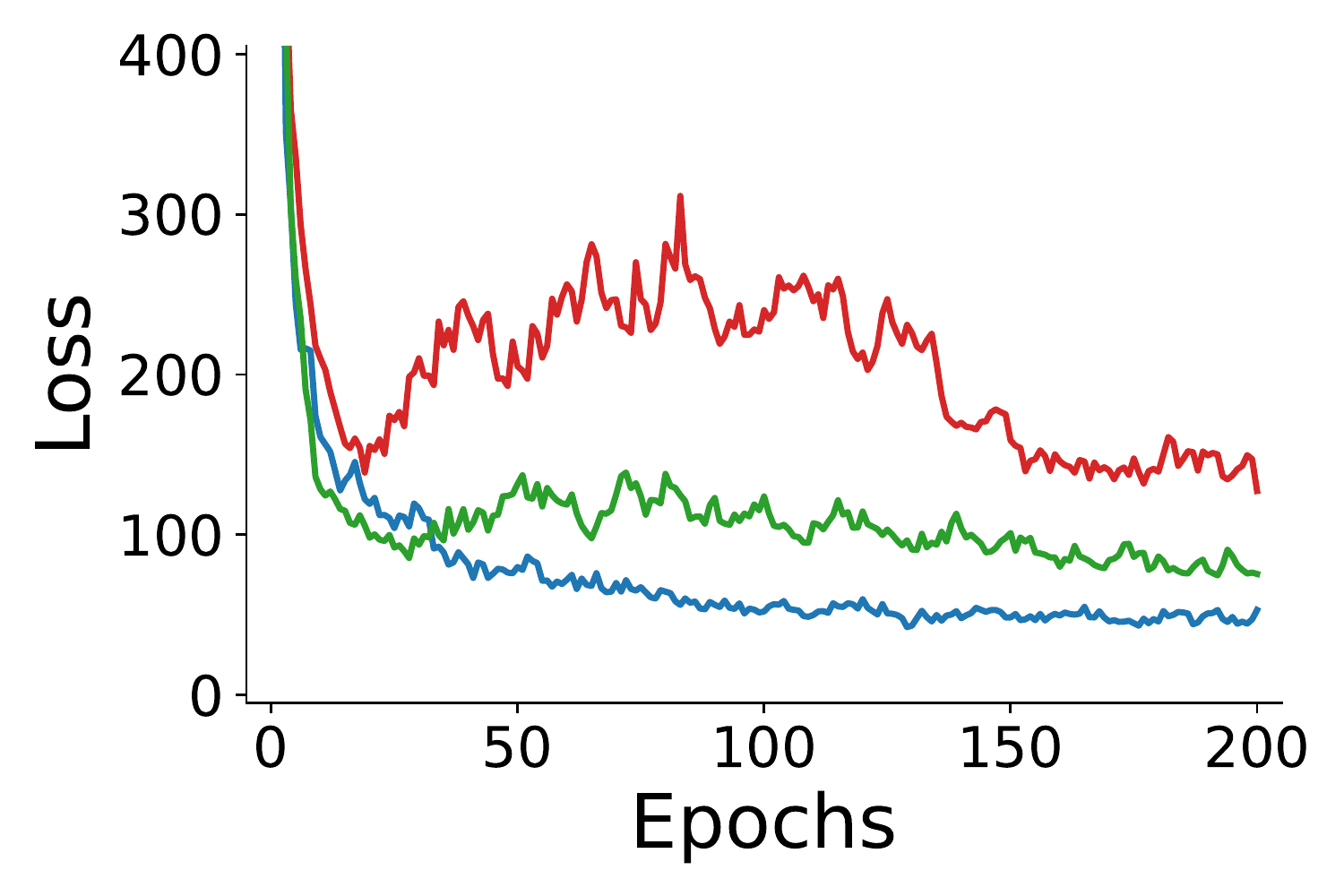}
    	\vspace{-15pt}
    	\caption{\small{\textsf{MovieLens 1M}}}
	\end{subfigure}
	\hfill
	\begin{subfigure}[b]{.475\linewidth}
    	\includegraphics[width=1.0\linewidth]{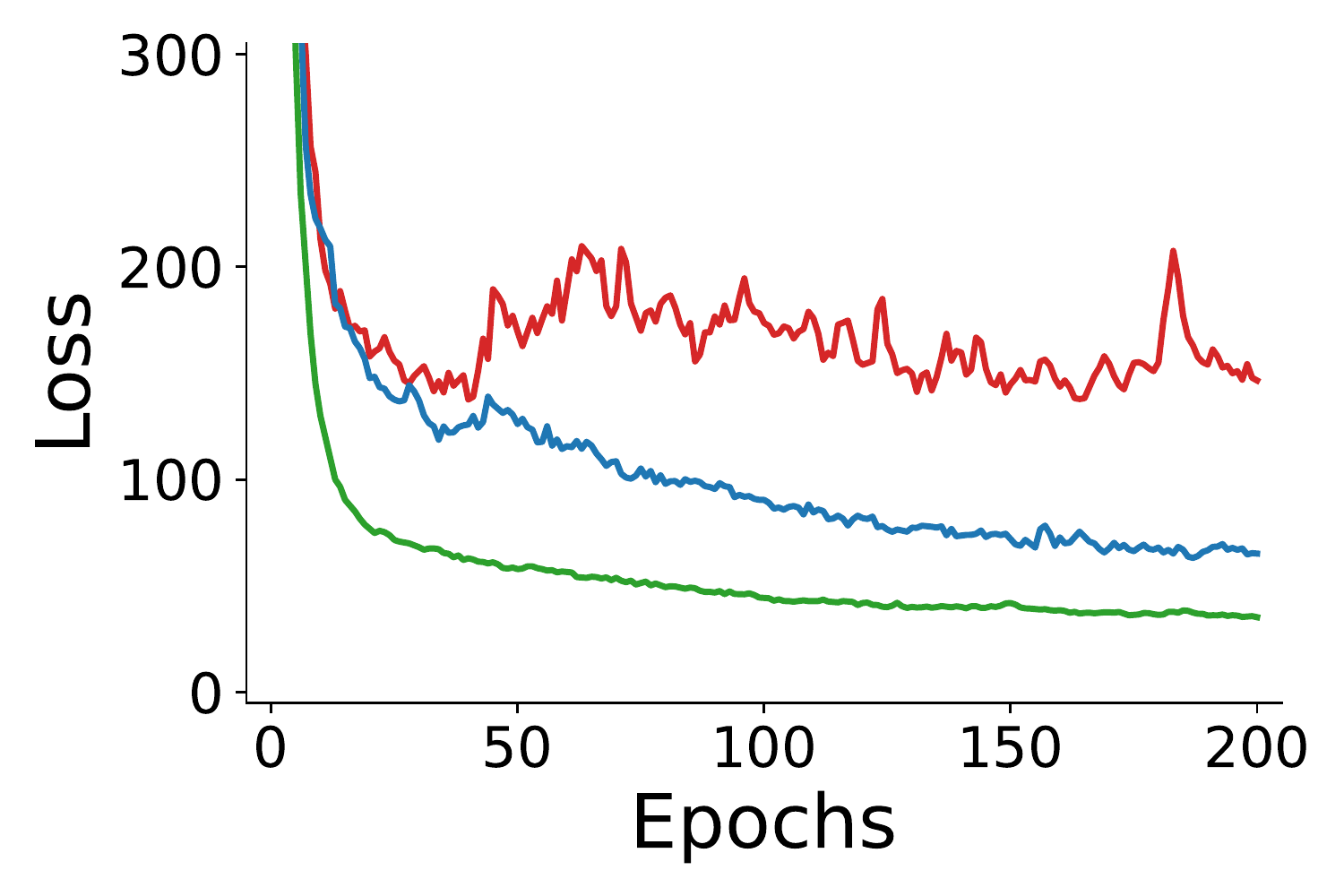}
    	\vspace{-15pt}
    	\caption{\small{\textsf{Yelp}}}
	\end{subfigure}
	\caption{The joint training scheme in \methodq facilitates and stabilizes optimization in \textsf{MovieLens 10M} and \textsf{Yelp}.  \label{fig:loss}}
\end{figure}

\smallsection{Effects of Joint Training:}
We analyze the effects of the joint training scheme of \methodq by comparing \methodbq ($\lambda=0$) and \methodq ($\lambda\geq 0$) and to this end, we measure the relative MSE defined as:
\begin{equation}\label{eq:relmse:joint}
    \frac{\text{MSE of \methodq}}{\text{MSE of \methodbq}}
\end{equation}
of each dataset.
Figure~\ref{fig:effect:joint} shows that \methodq is superior compared to \methodbq in most datasets indicating that jointly training the reconstruction boxes with the original ones leads to accurate boxes.
As summarized in Table~\ref{tab:effects:methodq}, joint training reduces the average MSEs on the considered datasets, by up to $44\%$, together with the box quantization scheme.
For example, together with the box quantization scheme, joint training reduces estimation error by $64\%$ and $38\%$ for the Jaccard Index on \textsf{Gplus} and for the Overlap Coefficient on \textsf{Netflix}, respectively.
These results imply that learning quantized boxes simultaneously with the original boxes improves the quality of the quantization and thus its effectiveness.
To further analyze these results, we investigate how the loss decreases with respect to $\lambda$.
In Figure~\ref{fig:loss}, we observe that training the reconstructed boxes alone ($\lambda=0$) is unstable, and learning the original boxes together ($\lambda>0$) helps not only stabilize but also facilitate the optimization.

\begin{table}[t]
	\begin{center}
		\caption{\label{tab:effects:methodq}
		The proposed schemes: box quantization and joint training in \methodq incrementally improves the accuracy (in terms of MSE) averaged over all considered datasets.
		}
		\scalebox{0.82}{
			\begin{tabular}{c|cccc}
				\toprule
				\textbf{Method} & \textbf{OC} & \textbf{CS} & \textbf{JI} & \textbf{DI}\\
				\midrule
				\methodpq & 0.0129 & 0.0028 & 0.0012 & 0.0023\\
				\methodbq & 0.0106 (\textcolor{myred}{-17\%}) & 0.0023 (\textcolor{myred}{-17\%}) & 0.0009 (\textcolor{myred}{-26\%}) & 0.0019 (\textcolor{myred}{-17\%})\\
				\methodq & \textbf{0.0077} (\textcolor{myred}{\textbf{-40\%}}) & \textbf{0.0016} (\textcolor{myred}{\textbf{-44\%}}) & \textbf{0.0007} (\textcolor{myred}{\textbf{-41\%}}) & \textbf{0.0013} (\textcolor{myred}{\textbf{-42\%}})\\
				\bottomrule %
			\end{tabular}}
	\end{center}
	\vspace{-3mm}
\end{table}

\smallsection{Effects of Boxes:}
To confirm the effectiveness of using boxes for representing sets, we consider \methodo, which is also a region-based geometric embedding method:
\begin{itemize}[leftmargin=*]
    \item \textbf{\methodo:} A set $s$ is represented as a $d$-dimensional vector $\mathrm{z}_s\in\mathbb{R}_{+}^{d}$ whose volume is computed as $\mathbb{V}(\mathrm{z}_s)=\exp(- \sum_i \mathrm{z}_{s}[i])$. 
    The volume of the intersection and the union of two representations $\mathrm{z}_{s}$ and $\mathrm{z}_{s'}$ of sets $s$ and $s'$, respectively, are:
    \begin{align*}
        &\mathbb{V}(\mathrm{z}_{s} \wedge \mathrm{z}_{s'}) = \exp \left(- \textstyle\sum\nolimits_i \max(\mathrm{z}_{s}[i], \mathrm{z}_{s'}[i])\right), \\
        &\mathbb{V}(\mathrm{z}_{s} \vee \mathrm{z}_{s'}) = \exp \left(- \textstyle\sum\nolimits_i \min(\mathrm{z}_{s}[i], \mathrm{z}_{s'}[i])\right),
    \end{align*}
    respectively.
    The encoding cost is $32d|\mathcal{S}|$ bits.
\end{itemize}
We set the dimensions for \methodo and \method to $8$ and $4$, respectively, so that their encoding costs are the same.
In Table~\ref{tab:effects:order}, we compare \method with \methodo in terms of the average MSE on the considered datasets for each measure.
\method yields more accurate representations than \methodo, implying the effectiveness of boxes to represent sets for similarity preservation.
For example, \method achieved $62\%$ lower average MSE than \methodo in preserving the Overlap Coefficient.

\begin{table}[t]
	\begin{center}
		\caption{\label{tab:effects:order}
		\method yields smaller MSE on average in the considered datasets than \methodo.
		}
		\scalebox{0.8}{
			\begin{tabular}{c|cccc}
				\toprule
				\textbf{Method} & \textbf{OC} & \textbf{CS} & \textbf{JI} & \textbf{DI}\\
				\midrule
				\methodo & 0.0320 & 0.0033 & 0.0008 & 0.0027\\
				\method & \textbf{0.0121} (\textcolor{myred}{\textbf{-62\%}}) & \textbf{0.0028} (\textcolor{myred}{\textbf{-14\%}}) & \textbf{0.0006} (\textcolor{myred}{\textbf{-22\%}}) & \textbf{0.0022} (\textcolor{myred}{\textbf{-17\%}})\\
				\bottomrule %
			\end{tabular}}
	\end{center}
\end{table}

\subsection{Q3. Effects of Parameters}\label{sec:experiments:parameters}
\begin{figure}[t]
	\centering
	\begin{subfigure}[b]{.475\linewidth}
    	\includegraphics[width=1.0\linewidth]{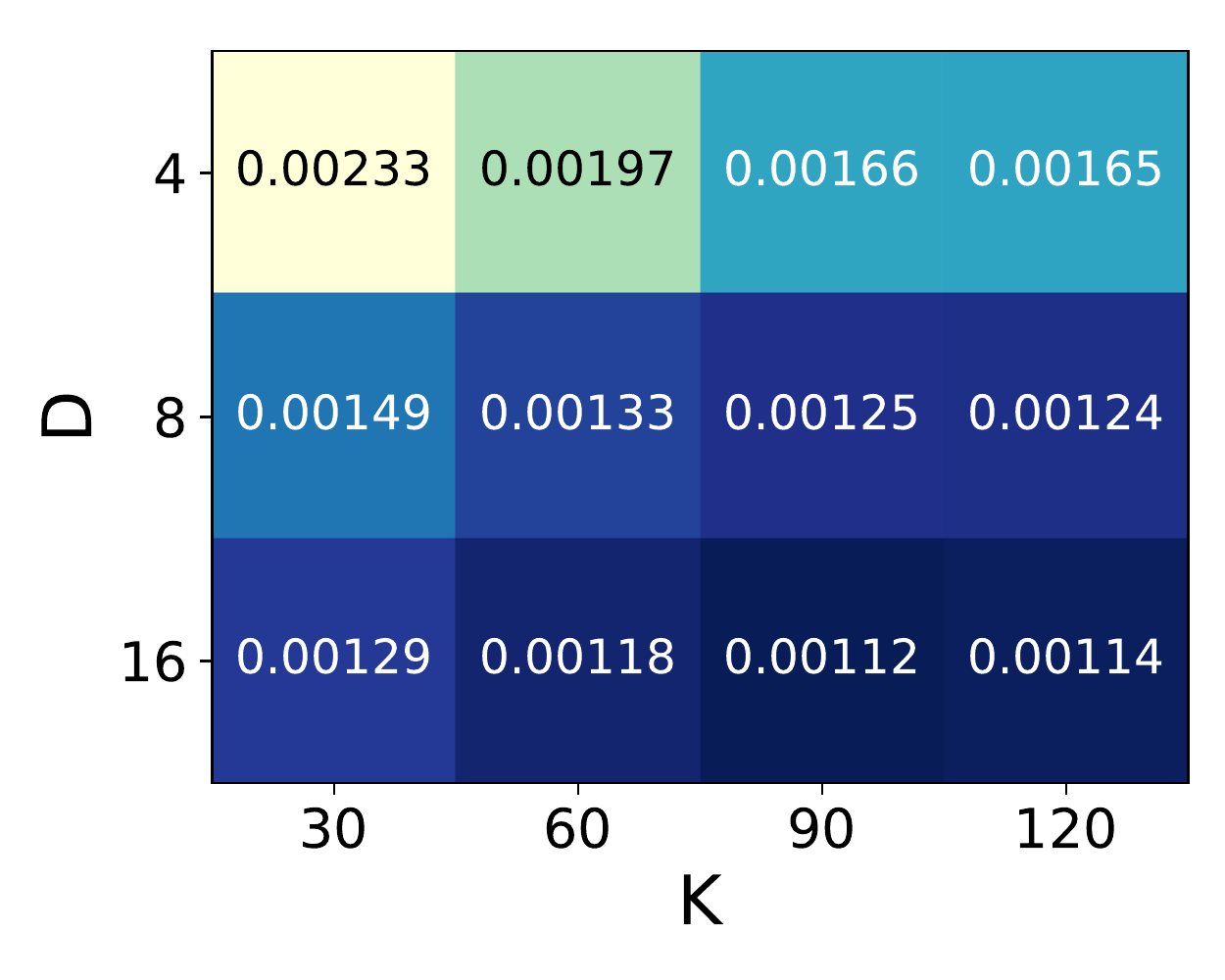}
    	\vspace{-15pt}
    	\caption{\textsf{MovieLens 1M}}
	\end{subfigure}
	\hfill
	\begin{subfigure}[b]{.475\linewidth}
    	\includegraphics[width=1.0\linewidth]{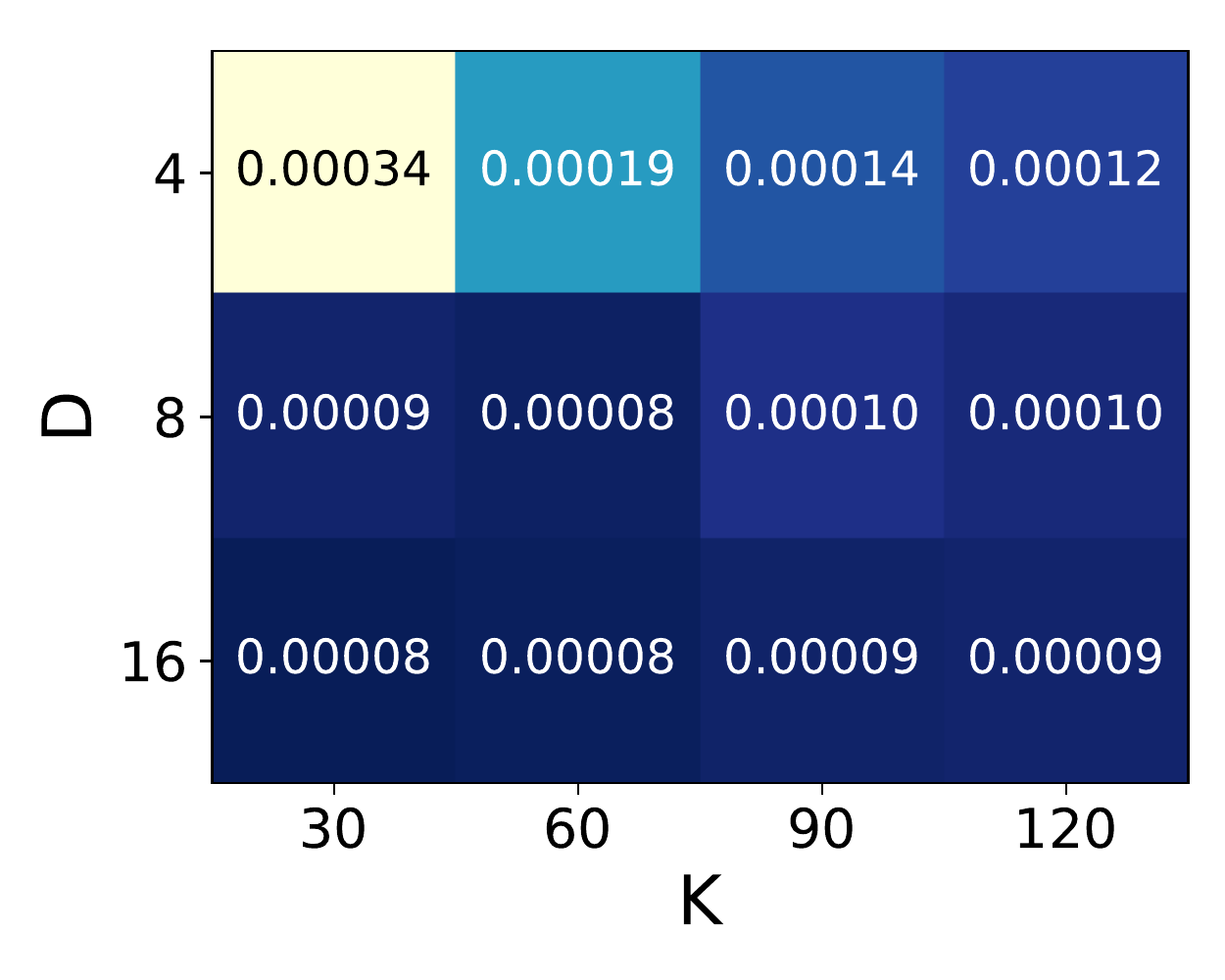}
    	\vspace{-15pt}
    	\caption{\textsf{Yelp}}
	\end{subfigure}
	\caption{Effects of $K$ and $D$ in \methodq on the approximation of the Jaccard Indices in \textsf{MovieLens 10M} and \textsf{Yelp}.\label{fig:kd}}
\end{figure}

We analyze how parameters of \methodq affect the embedding quality of the set representations. 
First, the number of subspaces ($D$) and the number of key boxes in each subspace ($K$) are the key parameters that control the encoding cost of \methodq. 
In Figure~\ref{fig:kd}, we investigate how the performance of \methodq depends on $D$ and $K$ values while fixing $d$ to $32$.
Typically, the accuracy improves as $D$ and $K$ increase at the expense of extra encoding cost. 
In addition, its performance is affected more heavily by $D$ than by $K$.

\begin{figure}[t]
    \centering
    \includegraphics[width=0.485\textwidth]{FIG/encoding_cost/encoding_cost_legend.pdf}\\
    \begin{subfigure}[b]{.475\linewidth}
        \includegraphics[width=1.0\linewidth]{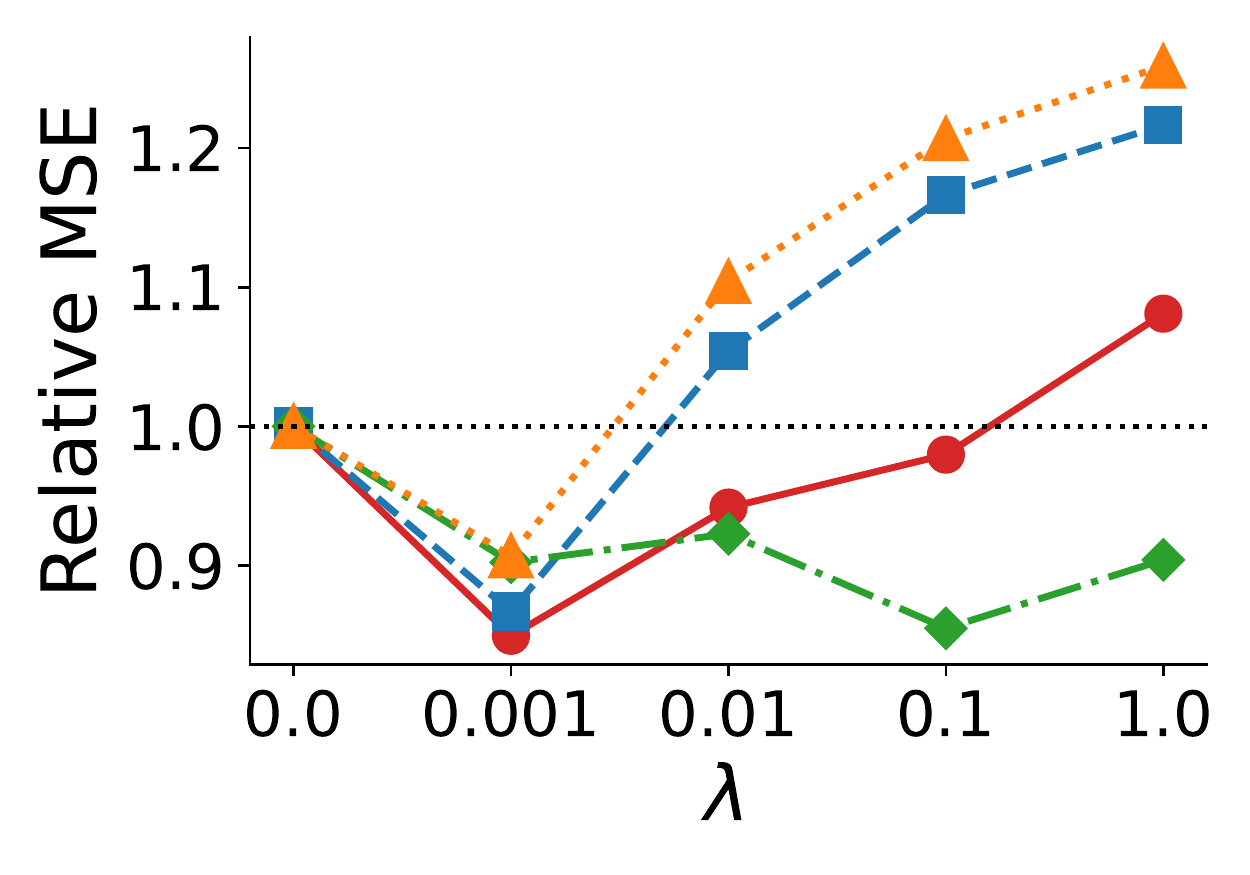}
        \vspace{-15pt}
        \caption{\textsf{MovieLens1M}}
    \end{subfigure}
    \hfill
    \begin{subfigure}[b]{.475\linewidth}
        \includegraphics[width=1.0\linewidth]{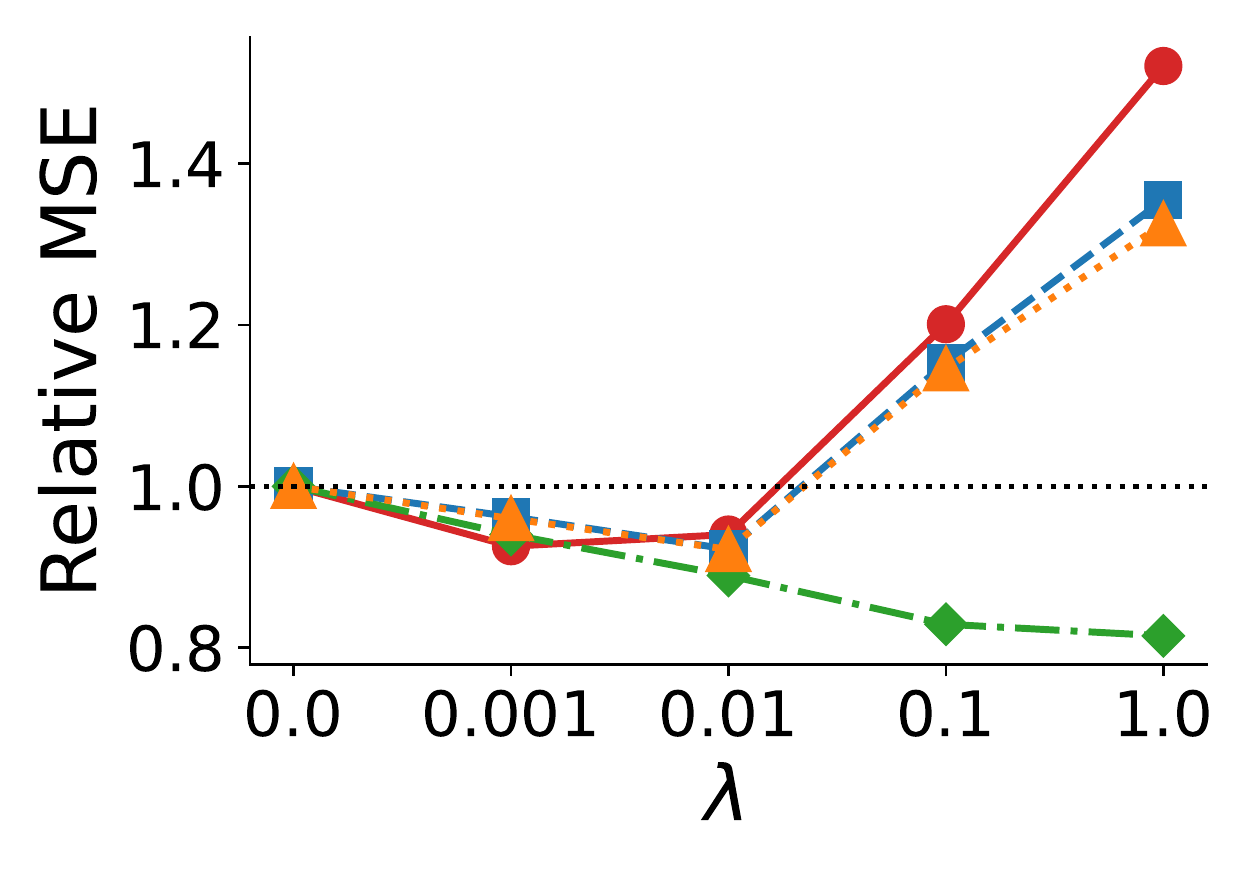}
        \vspace{-15pt}
        \caption{\textsf{Yelp}}
    \end{subfigure}
    \caption{Effects of $\lambda$ in \methodq.\label{fig:lambda}}
\end{figure}

In Figure~\ref{fig:lambda}, we observe how the coefficient $\lambda$ in Eq.~\eqref{eq:methodq_loss} affects the accuracy of \methodq.
To this end, we measure relative MSE (relative to MSE when $\lambda=0$) with different $\lambda$ values.
As shown in Figure~\ref{fig:lambda}, joint training is beneficial, but overemphasizing the joint views sometimes prevents \methodq from learning meaningful reconstructed boxes.

\begin{figure}[t]
    \centering
    \includegraphics[width=0.485\textwidth]{FIG/encoding_cost/encoding_cost_legend.pdf}\\
    \begin{subfigure}[b]{.475\linewidth}
        \includegraphics[width=1.0\linewidth]{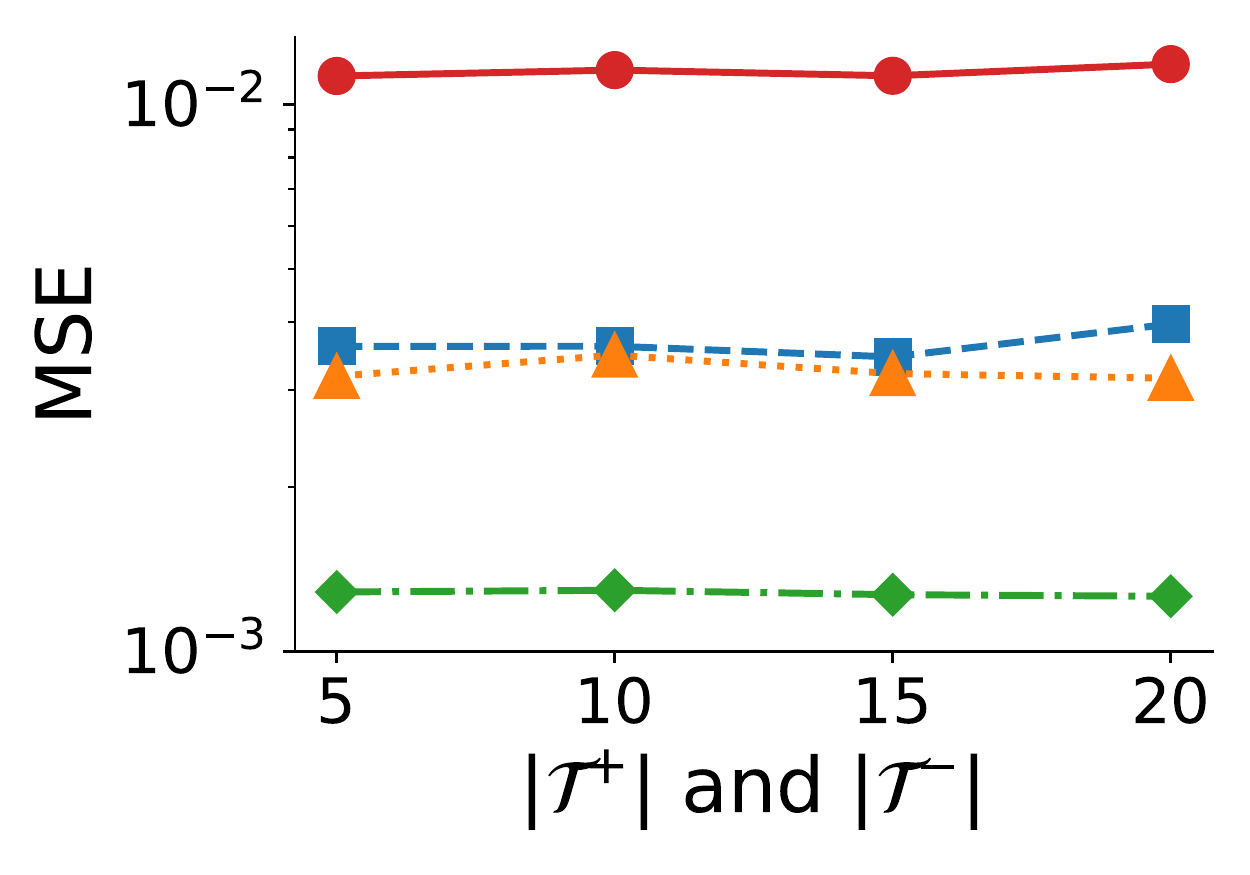}
        \vspace{-15pt}
        \caption{\textsf{MovieLens1M}}
    \end{subfigure}
    \hfill
    \begin{subfigure}[b]{.475\linewidth}
        \includegraphics[width=1.0\linewidth]{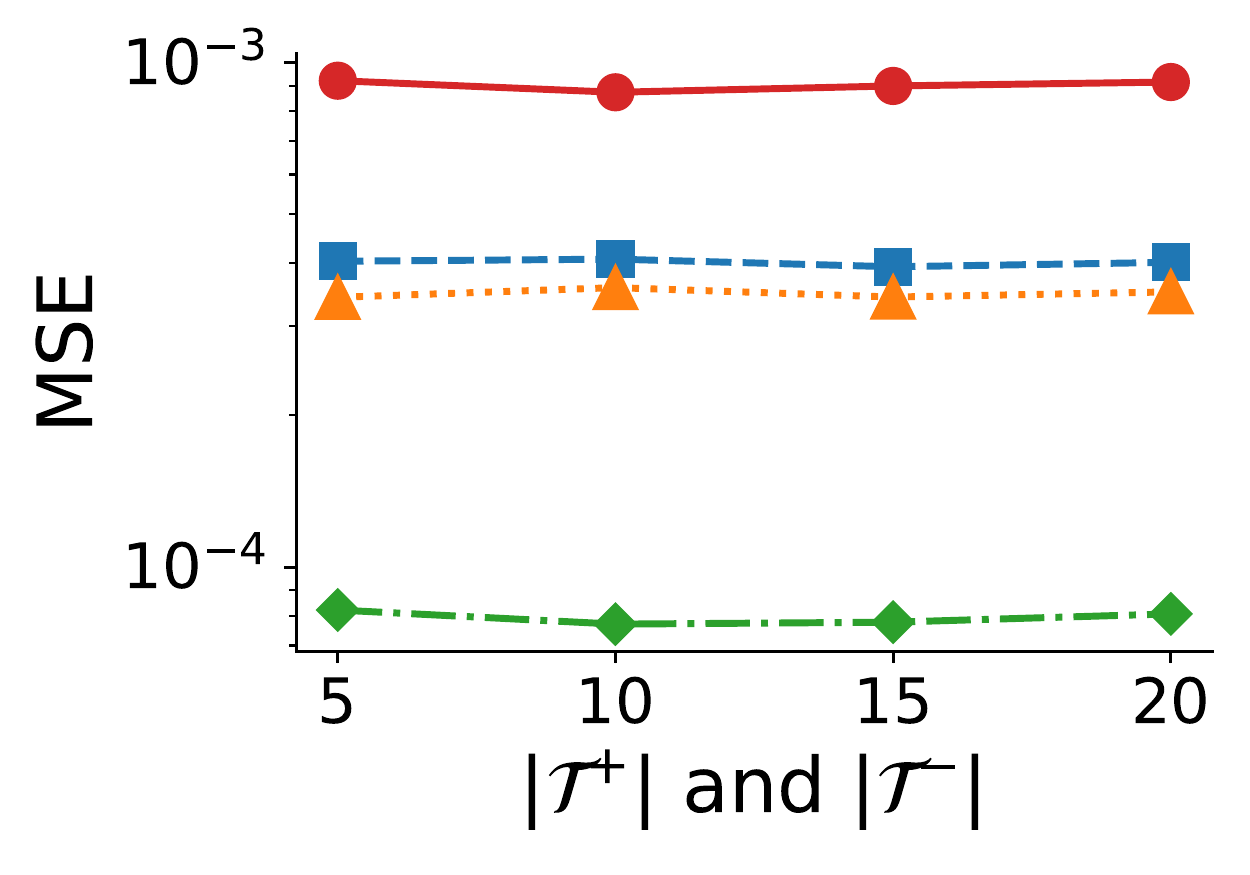}
        \vspace{-15pt}
        \caption{\textsf{Yelp}}
    \end{subfigure}
    \caption{Effects of $|\mathcal{T}^{+}|$ and $|\mathcal{T}^{-}|$ in \methodq.\label{fig:T}}
\end{figure}

In Figure~\ref{fig:T}, we observe the effects of the number of training samples, $|\mathcal{T}^{+}|$ and $|\mathcal{T}^{-}|$, in \methodq.
We can see that the accuracy is robust to the parameters, and thus using only a small number of samples for training is enough (we consistently use $|\mathcal{T}^{+}|=|\mathcal{T}^{-}|=10$ in all experiments).


	\section{Discussions}
	\label{sec:discussions}
	To further support the effectiveness of the proposed methods, \method and \methodq, we analyze the properties of boxes and other representation methods and find their relations with sets.
To this end, we review the following questions for the methods used in Section~\ref{sec:experiments},
as summarized in Table~\ref{tab:properties}:


\noindent
\textbf{RQ1. Are basic set properties supported?}

\noindent
\textbf{A1.} Boxes naturally satisfy six representative set properties, which are listed in Table~\ref{tab:box_properties}.
These properties are also met by \rh and \methodo, but not by the vector-based methods \vect and \vectmlp, since they do not contain information on the set itself (e.g., set sizes).

\noindent
\textbf{RQ2. Are sets of any-size representable?}

\noindent
\textbf{A2.} 
In \method and \methodq, boxes of various volumes can be learned by adjusting their offsets, and thus sets of any sizes are accurately learnable. 
So does the \methodo, by controlling the L1 norm of the vector. 
However, \rh inevitably suffers from information loss for sets larger than $d$ (see Figure~\ref{fig:crown} in Section~\ref{sec:prelim}). 
The vector-based methods have no limitations regarding set sizes.

\noindent
\textbf{RQ3. Are representations expressive enough?}

\noindent
\textbf{R3.}
Boxes of diverse shapes and sizes can be located anywhere in the Euclidean latent space by controlling their centers and offsets.
This property provides boxes with expressiveness, enabling them to capture complex relations with other boxes.
In \methodo, a single nonnegative vector is learned to control the volume of the region. 
This nonnegativity limits the expressiveness of the embeddings.
\rh suffers from hash collisions, and thus different sets can be represented as the same binary vector, which causes considerable information loss if $d$ is not large enough.
Despite their wide usage in various fields, empirically, \vect and \vectmlp have limited power to accurately preserve similarities between sets.
In particular, set embeddings obtained from them for a specific measure are not extensible to estimate other measures.

\begin{table}[t]
	\begin{center}
		\caption{\label{tab:properties}
		Properties of the considered methods regarding Q1, Q2, and Q3 in Section~\ref{sec:discussions}.
		}
		\scalebox{0.975}{
			\begin{tabular}{c|ccccccccc}
				\toprule
				\textbf{Method} && \textbf{Q1} &&& \textbf{Q2} &&& \textbf{Q3} & \\
				\midrule
				\rh && \cmark &&& \xmark &&& \xmark &\\
				\vect \& \vectmlp && \xmark &&& \cmark &&& \xmark &\\
				\methodo && \cmark &&& \cmark &&& \xmark &\\
				\midrule
				\method \& \methodq && \cmark &&& \cmark &&& \cmark &\\
				\bottomrule 
			\end{tabular}}
	\end{center}
\end{table}

	\section{Conclusions}
	\label{sec:summary}
	In this work, we propose \method, an effective representation learning method for preserving similarities between sets.
Thanks to the unique geometric properties of boxes, \method accurately preserves various similarities without assumptions about measures.
Additionally, we develop \methodq, which is equipped with novel box quantization and joint training schemes.
Our empirical results support that \methodq has the following strengths over its competitors:
\begin{itemize}[leftmargin=*]
    \item \textbf{Accurate:} \methodq yields up to $40.8\times$ smaller estimation error than competitors, requiring smaller encoding costs.
    \item \textbf{Concise:} \methodq requires up to $96.8\times$ smaller 
    encoding costs
    to achieve the same accuracy of the competitors.
    \item \textbf{Versatile:} \methodq is free from assumptions about similarity measures to be preserved.
\end{itemize}
For \textbf{reproducibility}, the code and data are available at \url{https://github.com/geon0325/Set2Box}.

 \begin{table}[t]
    	\begin{center}
    		\caption{\label{tab:box_properties}
    		Boxes satisfy various set properties.
    		}
    		\scalebox{0.90}{
    			\begin{tabular}{l|c|c}
    				\toprule
    				\textbf{Property} & \multicolumn{2}{c}{\textbf{Properties Satisfied by Boxes }} \\
    				\midrule
    				1. Transitivity Law & \multicolumn{2}{c}{$\mathrm{B}_X \subseteq \mathrm{B}_Y,\;\mathrm{B}_Y \subseteq \mathrm{B}_Z \rightarrow \mathrm{B}_X \subseteq \mathrm{B}_Z$}\\
    				\midrule
    				2. Idempotent Law & $\mathrm{B}_X \cup \mathrm{B}_X = \mathrm{B}_X$ & $\mathrm{B}_X \cap \mathrm{B}_X = \mathrm{B}_X$ \\
    				\midrule
    				3. Commutative Law & $\mathrm{B}_X \cup \mathrm{B}_Y = \mathrm{B}_Y \cup \mathrm{B}_X$ 
    				& $\mathrm{B}_X \cap \mathrm{B}_Y = \mathrm{B}_Y \cap \mathrm{B}_X$\\
    				\midrule
    				\multirow{2}{*}{4. Associative Law} & \multicolumn{2}{c}{$\mathrm{B}_X \cup (\mathrm{B}_Y \cup \mathrm{B}_Z) = (\mathrm{B}_X\cup \mathrm{B}_Y)\cup \mathrm{B}_Z$}\\
    				& \multicolumn{2}{c}{$\mathrm{B}_X \cap (\mathrm{B}_Y \cap \mathrm{B}_Z) = (\mathrm{B}_X\cap \mathrm{B}_Y)\cap \mathrm{B}_Z$}\\
    				\midrule
    				5. Absorption Law & $\mathrm{B}_X \cup (\mathrm{B}_X \cap \mathrm{B}_Y) = \mathrm{B}_X$
    				& $\mathrm{B}_X \cap (\mathrm{B}_X \cup \mathrm{B}_Y) = \mathrm{B}_X$\\
    				\midrule
    				\multirow{2}{*}{6. Distributive Law} & \multicolumn{2}{c}{$\mathrm{B}_X\cap (\mathrm{B}_Y \cup \mathrm{B}_Z) = (\mathrm{B}_X \cap \mathrm{B}_Y) \cup (\mathrm{B}_X \cap \mathrm{B}_Z)$}\\
    				& \multicolumn{2}{c}{$\mathrm{B}_X\cup (\mathrm{B}_Y \cap \mathrm{B}_Z) = (\mathrm{B}_X \cup \mathrm{B}_Y)\cap (\mathrm{B}_X \cup \mathrm{B}_Z)$}\\
    				\bottomrule 
    			\end{tabular}}
    	\end{center}
    \end{table}

	\bibliographystyle{IEEEtran}
	\bibliography{BIB/ref}
	\balance


\end{document}